\documentclass[11pt]{iopart}

\usepackage{epsf}
\usepackage{longtable}
\usepackage{hyperref}
\usepackage{graphics,epsfig,placeins,wrapfig}
\expandafter\let\csname equation*\endcsname\relax

\expandafter\let\csname endequation*\endcsname\relax
\usepackage{amsmath,amssymb,calc,amsfonts}
\usepackage[usenames]{color}
\usepackage{soul}
\usepackage{amsfonts}
\usepackage[mathscr]{euscript}
\usepackage{subcaption}

\DeclareMathAlphabet\mathbfcal{OMS}{cmsy}{b}{n}

\def\eadnew#1#2{\address{#2 E-mail: \mailto{#1}}} 

\DeclareUnicodeCharacter{2212}{-}

\begin{document}

\title{Synchrotron emitting Komissarov torus around naked singularities}

\author{German D. Prada-Méndez $^{1}$, 
F. D. Lora-Clavijo $^{1}$ \footnote{Corresponding author},
J. M. Velásquez-Cadavid $^{1}$}
\address{$^1$  Escuela de F\'isica, Universidad Industrial de Santander, A. A. 678,
Bucaramanga 680002, Colombia.}

\eadnew{fadulora@uis.edu.co}{}

\pacs{04.20.-q; 04.40.-b; 47.10.-g}

\begin{abstract}
From a theoretical perspective, matter accretion processes around compact objects are highly relevant as they serve as a natural laboratory to test general relativity in the strong field regime. This enables us to validate fundamental concepts such as the no-hair theorem, the cosmic censorship hypothesis, and the existence of alternative solutions to Einstein's equations that mimic the effects of black holes. In this study, we analyze the emission spectra of geometrically thick accretion disks,  referred to as Polish doughnuts, around naked singularities described by the $q$-metric. To begin, we revisit the construction of equilibrium configurations of magnetized tori in this spacetime and evaluate the role of the deformation parameter over these configurations. Once we have systematically studied the disks in this spacetime, we use the \texttt{OSIRIS} code to perform a backward ray-tracing method, resulting in the first simulations of the intensity map and emission profiles of magnetized tori within  this metric. Furthermore, we validate the effect of both the quadrupole moment and the angular momentum on observable quantities such as flux and intensity for optically thin and thick disks, since for values of $ q < 0$, which correspond to objects with prolate deformation, and which in turn, are constructed with higher values of angular momentum, the emission spectrum exhibits higher intensity than that obtained for Schwarzschild's spacetime. Hence, we find a first differential feature that distinguishes tori formed around naked singularities from those around static black holes.
\end{abstract}

\section{Introduction}

Compact objects are one of the most appealing predictions of general relativity, as their existence plays a significant  role in explaining diverse astrophysical phenomena such as gravitational lensing and the shadow cast due to the radiation emitted by the accretion disk around a compact object, among others. The first observational evidence linked with these objects came from studying high-energy phenomena such as quasars \cite{rees1978quasars} and the production of relativistic jets  \cite{blandford1979relativistic}. Likewise, these have been associated with quasi-periodic oscillations and the orbits of various objects in the cosmos \cite{abuter2020detection}. Further, it is presumed that these structures are present in galactic centres. On the other hand, it has been postulated that the accretion of matter onto compact objects is one of the most important ways of energy production in the universe, converting gravitational energy into thermal and electromagnetic radiation. On top of that, it is relevant in X-ray production processes.    

Recently, additional observations associated with the gravitational effects evidenced on light and matter in the vicinity of compact objects have been made. Firstly, measurements of the trajectories followed by stars orbiting these structures have been registered. The data relating to these measurements were collected by the GRAVITY collaboration \cite{abuter2018detection}. Of particular interest are the S-stars, the closest star cluster to the supermassive black hole Sagittarius A* located at the centre of the Milky Way. Among this group, the S2 star is of particular interest due to its relative brightness and close trajectory near the compact object, which allows for a better constriction of the spacetime in this vicinity. 

On the other hand, observations recorded by the Event Horizon Telescope (EHT) collaboration \cite{akiyama2019first} resulted in the first-ever image of the shadow \footnote[2]{A term coined by \cite{falcke1999viewing}.} produced by the supermassive black hole candidate at the centre of the M87 galaxy \cite{collaboration2019first}, which is consistent with a Kerr black hole with a mass of $(6,5 \pm 0,7)\times 10^9$ $M_\odot$  and a rotation parameter $ a = 0.90 \pm 0.05$. Additionally, the EHT collaboration obtained the most recent observation of the shadow of the compact object in the centre of the Milky Way \cite{akiyama2022first}, which is consistent with a black hole of $10^6$ $M_\odot$.  These images are obtained from the emission of accretion disks formed around black hole candidates.

Moreover, it should be noted that among the various indirect observational evidence of compact objects, the most abundant in the universe are those related to the emission spectrum of the accretion disks that form around them \cite{ho1997search, richstone1998supermassive}. Therefore, to study these structures,  the emission profiles emitted by the system and its temporal variations are analyzed. A review of the radiative properties in thin disks is presented in \cite{shaikh2019can, liu2021thin, liu2022thin, gyulchev2020observational, gyulchev2021image, gyulchev2019image, harko2009can}. Furthermore, the usefulness of the analysis of the emission spectra for different applications has been studied, such as relativistic jets \cite{mckinney2012general}, the appearance of turbulence and its relationship with the accretion processes \cite{balbus1991powerful, balbus1998instability}, and the study of instabilities  \cite{balbus1991powerful, papaloizou1984dynamical, abramowicz1983runaway, bugli2018papaloizou}. 

Besides,  the recent measurements of the shadows of compact objects by the EHT have sparked a revolution, highlighting the importance of studying accretion disks. These have enabled us to make comparisons, for the first time, between theoretical models and indirect observations of compact objects. Firstly, there is the validation of the no-hair theorem, which states that the electromagnetic and gravitational fields of a stationary black hole are determined solely by its mass, charge, and intrinsic angular momentum \cite{bardeen1973four}. Likewise, the cosmic censorship hypothesis can be put to test,  which suggests that it is not possible to have a singularity that is not surrounded by an event horizon \cite{penrose2002golden}. Finally, we can examine the validity of alternative solutions to Einstein's equations, which deviate from the traditional black hole models and may be consistent with observations. These solutions are often referred to as "black hole mimickers" as they reproduce the various effects of a black hole in its vicinity\footnote[3]{These solutions are often referred to as "black hole mimickers" since they replicate the different effects of a black hole in its vicinity.}.    

Thus, the modelling of these structures has been a relevant theme, going from purely hydrodynamic models to more elaborate models\footnote[4]{For a detailed review, see \cite{abramowicz2013foundations}.}. Among all the different proposals, the so-called "Polish Doughnuts" stand out \cite{abramowicz1978relativistic}, these are characterized by being the simplest relativistic stationary configurations that describe an ideal fluid rotating with constant angular momentum around a naked singularity. Moreover, the Polish doughnuts model was extended to include the magnetic field effects, specifically with a toroidal distribution, around a rotating black hole \cite{komissarov2006magnetized}. This work has served as the base for the construction of geometrically thick disks with magnetic fields and auto-consistent with the magnetohydrodynamic equations. 

This model extension proved to be of great interest due to the role of the magnetic fields in the outward transport of angular momentum, induced by the magneto-rotational instability (MRI) \cite{zanotti2015zeipel}  and its effect on the calculation of the synchrotron radiation emissivity \cite{wardzinski2000thermal}. On top of that, within this framework, the emission spectrum of Sgr A* was studied. It was found that the toroidal magnetic field model can reproduce the spectral measurements in the millimetre range of the radio band, where the spectral peak of Sgr A* is located \cite{vincent2014magnetized}. 

Among the advances achieved through this model, the dynamic evolution of disks presented by \cite{montero2007dynamics} and the inclusion of the case with non-constant angular momentum distribution by \cite{gimeno2017magnetised} stand out. In addition, \cite{wielgus2015local} proposes a model with a power-law distribution of angular momentum to explore the growth of the MRI \cite{balbus1991powerful}. Furthermore, the dynamic evolution of these configurations is addressed by \cite{fragile2017decay}, focusing on the decay of the magnetization. Likewise,  the recent work of \cite{pimentel2018analytic} is highlighted, in which the effect of magnetic polarization is taken into account. This model has been extended recently to study the effect of a non-constant angular momentum distribution over magnetically polarized tori \cite{gimenosoler2023magnetised}. Similarly, other works have been carried out in this area to incorporate effects such as self-gravitation of the disk \cite{mach2019self}, and viscosity of the fluid \cite{lahiri2021stationary}.   

On account of recent observations and advancements in disk modelling,  new avenues of research have emerged, including a focus on constraining possible solutions to general relativity (GR) and other gravity theories \cite{shaikh2021constraining, psaltis2020gravitational, kumar2020black}. In general, it is assumed that most compact objects are described through the "traditional" models which describe black holes\footnote[5]{Mainly through the Schwarzschild and Kerr metrics}, and have been validated through observations; however, other models are solutions to Einstein's equations and are also consistent with current observations. Hence, these should not be discarded yet.  It is common to find in the literature studies on the observational properties of various compact objects based on models of the accretion disk spectrum \cite{boshkayev2020accretion, kovacs2010can, joshi2013distinguishing}.  

Therefore,  systematic comparisons of the theoretical shadows for M87 with the different models have been carried out, to identify restrictions on the parameter space of each of the solutions \cite{shaikh2021constraining} and to limit the alternative models to black holes. This process is possible because certain features of accretion disks depend on the spacetime geometry \cite{abramowicz2013foundations}. Hence, these structures are considered an ideal framework for the distinction between black holes and their mimickers.  

Additionally, alternative models cannot be ruled out in favour of the traditional ones since studies conducted on the latter still do not fully characterize the geometry of the space near the compact object. On the contrary, it has been shown that some of the mimickers are still plausible alternatives in light of observational data. \cite{bambhaniya2019timelike, bambhaniya2019timelike2, bambhaniya2021precession, bambhaniya2021shadows, nampalliwar2021modeling, jusufi2021black, saurabh2021imprints, joshi2020shadow, dey2019towards, lemos2008black}. In general, it has been demonstrated that any compact object with a photon sphere can cast shadows very similar to those of a black hole \cite{shaikh2019can}.

Among the alternatives, there are boson stars \cite{kaup1968klein, lora2010evolution, liebling2017dynamical, vincent2016imaging}, whose shadow is pretty similar to that of a Kerr black hole; gravastars \cite{mazur2004gravitational, sakai2014gravastar}, derived from the idea of Bose-Einstein condensate; solutions with dark matter \cite{boshkayev2020accretion, becerra2021hinting}, either completely formed nuclei of exotic matter or dark matter halos orbiting compact objects; regular black holes \cite{hayward2006formation}, which are characterized by the absence of a curvature singularity; Yukawa black holes \cite{cruz2021magnetized}, described by the  $f(R)$ gravity model; and finally, more exotic options, such as wormholes \cite{bambhaniya2022thin, gao2017traversable}.

Likewise, there are naked singularities, compact objects with a curvature singularity but without an event horizon. The existence of this type of solution would result in the violation of the cosmic censorship hypothesis \cite{christodoulou1984violation}. However, it has been shown that starting from physically reasonable conditions, it is possible to collapse a fluid in such a way that leads to an object of this type \cite{shapiro1991formation}. Moreover, this type of solution is relevant since it allows for studying gravitational phenomena in the ultra-strong regime, i.e., in the vicinity of a curvature singularity \cite{shaikh2019shadows}. Studies in this area have been carried out under different conditions and for a variety of spacetime metrics \cite{abdikamalov2019black} \cite{babar2017periodic} \cite{liu2018distinguishing} \cite{gyulchev2019image}.

In particular, it is of interest the spacetime described by the $q$-metric \cite{quevedo2011mass}, as it is the simplest static and axisymmetric solution to Einstein's equations with non-vanishing mass quadrupole \cite{quevedo2011mass}. This solution is obtained through a Zipoy-Vorhees transformation \cite{zipoy1966topology,voorhees1970static} of the Schwarzschild metric; such solutions display intriguing properties, including the emergence of repulsive gravity zones. This has been described in terms of the analysis of an effective potential \cite{quevedo2011mass}, and more recently through the study of the eigenvalues of the Riemann tensor of this spacetime \cite{arrieta2020shadows}.

Furthermore,  several studies have been conducted on the characteristics of the $q$-metric. For instance, the theoretical shadow of a naked singularity described by this metric has been simulated \cite{arrieta2020shadows,shaikh2021shadows}, as well as the Einstein ring and the gravitational lensing effects \cite{arrieta2020shadows}, geodesics  \cite{chowdhury2012circular}, quasi-normal modes \cite{allahyari2019quasinormal}, trajectories of charged particles \cite{benavides2019charged}, rotating particles  \cite{toshmatov2019spinning}, neutrino oscillations \cite{boshkayev2020neutrino}, and particle collisions, amid others. 

There are few references in the literature regarding the study of accretion disks around naked singularities described by this metric. For example, studies exist for the case of thin hydrodynamic disks using the Novikov-Thorne approach \cite{boshkayev2021luminosity, shaikh2021shadows}. On the other hand, recent work has been done on thick hydrodynamic disks such as that of \cite{memmen2021geometrically}, as well as extensions that incorporate a rotation \cite{faraji2022relativistic} parameter.

However, a crucial element to differentiating between black hole models and their mimickers is the identification of observable parameters that allow contrasting different theoretical frameworks in the light of observations. In this sense, computer simulations that allow for studying the radiation coming from the accretion disk around the compact object are relevant. This radiation consists of a series of photons emanating from the disk that escape the gravitational pull of the source. To this end, a backward ray-tracing method is used to reconstruct the null geodesics that characterize this radiation. Consequently, countless studies of plasma near compact objects have been carried out.  These include the first computational image of a spherical black hole surrounded by a thin accretion disk and the characterization of emission lines from the same type of structures \cite{luminet1979image}.    

Recently, other works in this area have been developed, now incorporating geometrically thick disks in the Kerr spacetime under both keplerian and sub-keplerian regimes \cite{pariev1998line, wu2007iron, velasquez2023synchrotron}. In addition, advances have been made using computational resources for non-conventional metrics that describe black hole mimickers. Specifically, the image of thin hydrodynamic disks around a naked singularity has been obtained  \cite{shaikh2019can, shaikh2021shadows}. Nevertheless, up to date, there is no characterization of the emission spectra or images of magnetized thick disks near a naked singularity. 

To conduct this study, there are several options available in the literature, including GeoKerr \cite{dexter2009fast}, Gyoto \cite{vincent2011gyoto}, and BHAC  \cite{porth2017black}, among others. However, we have decided to utilize our authorship-developed code, \texttt{OSIRIS} code \cite{velasquez2022osiris}, which employs the Hamiltonian formulation to calculate null geodesics. We aim to determine changes in observational parameters that allow us to distinguish between the $q$-metric spacetime and the Schwarzschild spacetime. To achieve this, we analyze the emission spectra of thick accretion disks with a toroidal magnetic field, which enables us to present the first simulations of emission lines and intensity maps for magnetized tori within this spacetime.       

The organization of this paper is as follows: in Sect. \ref{qmet} we provide a general description of the $q$-metric and its multipole moments. Then, in Sect. \ref{construc} we present the model for the construction of thick disks, based on the work of \cite{pimentel2018analytic}, and describe the criteria for obtaining steady-state configurations systematically. In Sect. \ref{ResI}, we present mass density plots for the tori and evaluate the impact of the quadrupole on them. Next, in Sect. \ref{sec:RTE} and \ref{sec:NS} we expose, respectively, the synchrotron radiation model and the numerical setup used for the simulations. Then, in Sect. \ref{ResII}, we present the emission profiles and intensity maps for optically thin and thick tori around a naked singularity described by the $q$-metric. Finally, in Sect. \ref{Conclusion}, we discuss the results and present the conclusions of this work. Throughout this paper, we use the (-,+,+,+) signature and geometrized units, for which $G = c = 1$.     

\section{$q$-metric spacetime} \label{qmet}

In general, static and axisymmetric spacetimes are described through Weyl's line element \cite{weyl1917theory}, whose simplest case is that of Schwarzschild, which only has a mass monopole moment. To generalize this solution, a Zipoy-Vorhees transformation \cite{zipoy1966topology, voorhees1970static} is performed, resulting in a family of solutions with non-vanishing quadrupole, characterized by a parameter $\gamma$\footnote[1]{Hence, this solution is often referred to in the literature as the $\gamma$-metric}. Furthermore, these solutions differ only in higher-order multipole moments and share a common problem: the analytical treatment of the metrics due to their complex mathematical structure \cite{quevedo1990multipole}.  

Therefore, a re-interpretation of the Zipoy-Vorhees transformation in terms of a new parameter $q$\footnote[2]{The parameter $\gamma$ can be redefined by the relation $\gamma = q + 1$} is favoured. The $q$-metric is a solution that describes the exterior gravitational field of a naked singularity with a non-zero quadrupole moment and, in addition, presents the advantage of being an exact solution that can be expressed analytically in a simple way \cite{quevedo2011mass}. 

The metric tensor of this spacetime can be expressed in spherical coordinates as 
\begin{eqnarray}
	\begin{aligned}
		\mathbf{g}=&\left(1-\frac{2 m}{r}\right)^{1+q} \mathrm{~d} t \otimes \mathrm{d} t-\left(1-\frac{2 m}{r}\right)^{-q}\left[\left(1+\frac{m^{2} \sin ^{2} \theta}{r^{2}-2 m r}\right)^{-q(2+q)}\right.\\
		&\left.\times\left(\mathrm{d} r \otimes \mathrm{d} r\left(1-\frac{2 m}{r}\right)^{-1}+r^{2} \mathrm{~d} \theta \otimes \mathrm{d} \theta\right)+r^{2} \sin ^{2} \theta \mathrm{d} \varphi \otimes \mathrm{d} \varphi\right] ,
	\end{aligned}
\label{ggmet}
\end{eqnarray}
where $q$ and $m$ are arbitrary parameters. For the limiting case $q=0$, this solution reduces to the spherically symmetric Schwarzschild metric. On the other hand, for the case where $m= 0$, regardless of the $q$ value, it can be rewritten as the Minkowski metric. Thus, each of the parameters is, at first glance, associated with the physical variables of the system. In the case of $m$, it seems related to a mass distribution, whereas $q$ is related to a deviation from spherical symmetry.  

This physical association of the parameters was corroborated via the calculation of the ADM mass \cite{boshkayev2016motion} and the multipole structure \cite{quevedo2011mass}. The latter is achieved through the invariant definition proposed by  \cite{geroch1970multipole, geroch1970multipole2}. In the case of mass multipoles 
\begin{eqnarray}
	M_{0}=(1+q) m, \quad M_{2}=-\frac{m^{3}}{3} q(1+q)(2+q).
	\label{multipole}
\end{eqnarray} 

Moreover, these two terms provide a comprehensive description of the object's multipole structure, as any higher-order moment can be expressed in terms of them. Equation (\ref{multipole}), reveals that the mass monopole corresponds to the ADM mass, while the quadrupole reflects the object's deformation. For prolate mass distributions, $M_{2}$ has positive values, whereas oblate configurations yield negative values. Besides, considering the relation between the monopole and the system's mass, the restriction $q > -1$ is imposed to maintain positive total mass values \cite{quevedo2011mass}.

On the other hand, the study of the Kretschmann scalar \cite{quevedo2011mass} $K = R_{\mu \nu \lambda \tau}R^{\mu \nu \lambda \tau}$ shows that this solution is characterized by the presence of a pair of curvature singularities, located at $r = 0$ and on the hypersurface $r = 2m$, for any values of $q$ and $m$ that differ from those which correspond to the previously mentioned special cases. In addition, two further singularities have been identified, which are always contained within the radius $r = 2m$ and only appear for $q \in (-1,-1+\sqrt{3/2})\backslash {0}$. It is worth noting that \cite{luongo2014characterizing} have shown that no additional horizon exists outside the aforementioned hypersurface, indicating that these singularities correspond to naked singularities.

This is of great physical interest on account of its apparent violation of Penrose cosmic censorship hypothesis \cite{penrose2002golden}, which states that every curvature singularity must be surrounded by an event horizon. Moreover, it has been shown that shadows do not imply black holes, even though not all naked singularities produce them \cite{shaikh2019shadows}, while this metric certainly describes one \cite{arrieta2020shadows}.  Finally, various tests suggest that the final state of a continuous gravitational collapse can either result in a black hole or a naked singularity depending on conditions of the collapse process \cite{joshi2007gravitational}.
 
Besides, in the literature, repulsive gravity zones have been reported in the vicinity of the gravitational source described by this metric. Initially, this effect was attributed to an effective mass, which was connected to a relativistic effective potential governing geodesic motion in this spacetime \cite{quevedo2011mass}. Although, this interpretation was coordinate-dependent. Subsequent studies characterized these effects using the eigenvalues of the Riemann tensor associated with this metric \cite{arrieta2020shadows}. This way, the dependence on specific coordinates is eliminated. On the contrary, the variation of the sign in one of the eigenvalues, which is intrinsically related to curvature, is linked to changes in the behaviour of the gravitational field.

Finally, it is important to highlight other works related to the properties of this kind of solution, which have examined the impact of the quadrupole on phenomena such as geodesic motion and the formation of equilibrium configurations in the form of thick and thin accretion disks around a compact object described by this metric \cite{boshkayev2016motion, chowdhury2012circular, kodama2003global, herrera2005non, herrera2000measuring, papadopoulos1981some, parnovskii1985type}.

\section{Magnetohydrodinamic equations} \label{construc}

To build accretion disk configurations, one starts with the fundamental conservation equations, which govern the behaviour of matter. These are the conservation of rest mass and energy-momentum   \cite{abramowicz2013foundations}. To describe the fluid, one can write a general form of the energy-momentum tensor, $ T_{\mu}^{\nu}$, taking into consideration phenomena such as viscosity and polarization, among others. Then, the diverse accretion disk models are obtained as solutions to the conservation equations for particular choices of $ T_{\mu}^{\nu}$. 

For the case of thick magnetized accretion disks\footnote{Also known as magnetized Polish Doughnuts}, \cite{komissarov2006magnetized} extended the Polish Doughnut model. This model has considered the simplest analytical form in the sense that it uses the test fluid approximation, in which the self-gravity effects of the accreting fluid are neglected, and the energy-momentum tensor only includes the contribution of a perfect fluid \cite{abramowicz2013foundations}. As a result of this extension, another term representing the effects of the magnetic field is taken into account, and therefore, the ideal MHD equations are to be solved. These are not more than the previously mentioned conservation equations accompanied by the relevant Maxwell equations  \cite{komissarov2006magnetized}.
 
In particular, for this work, we assume a series of conditions for the description of the disks. Firstly, we have that the spacetime is stationary and axisymmetric, such that  $\partial_t g_{\mu \nu}=\partial_\phi g_{\mu \nu}=0$, where  ($t$, $\phi$, $r$, $\theta$) are the coordinates. Likewise, we assume that the flux is also stationary and axisymmetric, meaning that any physical parameter of the fluid satisfies $\partial_tf=\partial_\phi f=0$. Moreover, the flux is considered to be a pure rotation around the compact object, thus $u^{r}=u^{\theta}=0$. Finally, we have a completely azimuthal magnetic field (toroidal distribution), thus $b^{r}=b^{\theta}=0$.  
 
By applying these conditions, we found that the baryonic number conservation and the Maxwell equations are identically satisfied. Then, the behavior of the fluid is described by the energy-momentum conservation equation with the given form of the energy-momentum tensor \cite{komissarov2006magnetized}. After extensive mathematical development, we can obtain the relativistic Euler equations
\begin{eqnarray}
	\partial_\nu\left(\ln \left|u_t\right|\right)-\frac{\Omega}{1-l \Omega} \partial_\nu l+\frac{\partial_\nu P}{w}+\frac{\partial_\nu\left( \mathcal{L} P_m\right)}{\mathcal{L} w}=0,
	\label{eulerdef}
\end{eqnarray}
where $w$ is the enthalpy density, $P$ is the hydrodynamic pressure,  $b^2 = b^\mu b_{\mu}$ and $g^{\mu\nu}$ are the components of the metric tensor. Also, the magnetic pressure is defined as $P_{m} = b^2/2$ and $\mathcal{L}=g_{t \phi}^{2}-g_{t t} g_{\phi \phi}$, whereas the angular velocity ($\Omega$) and the angular momentum ($l$) are described by the following expressions
\begin{eqnarray}
	\Omega=\frac{u^{\phi}}{u^{t}}=-\frac{g_{\phi t}+l_{t t}}{g_{\phi \phi}+l g_{t \phi}},
	\label{vang}
\end{eqnarray}
\begin{eqnarray}
	l=-\frac{u_{\phi}}{u_{t}}=-\frac{g_{\phi t}+\Omega g_{\phi \phi}}{g_{t t}+\Omega g_{t \phi}}.
	\label{momang}
\end{eqnarray}

Once these definitions are incorporated, we can rewrite the Euler relativistic equations (\ref{eulerdef}) in the form of an exact differential; this is done following \cite{komissarov2006magnetized}. To achieve this, we need to specify a couple more features of the fluid. In particular, we assume that the fluid follows a barotropic equation of state $w = w(P)$ and that $\Omega = \Omega (l)$, thus the surfaces of equal $\Omega$ and $l$ coincide \cite{abramowicz1978relativistic}. Then, the integrability conditions \cite{komissarov2006magnetized} are applied. Thus, one can obtain an equation to characterize the fluid behavior.
     
Moreover, we introduce the relativistic effective potential $W$ \cite{abramowicz1978relativistic}. This is the generalization of the Newtonian relativistic potential. In the case of a disk, one haves that on its inner edge  $P = P_m = 0$, $u_t = u_{t_{in}}$ y $l = l_{in}$. As a result, an integrated form of equation  (\ref{eulerdef}) in terms of the relativistic effective potential $W$, where  $W_{in}$ is a constant that denotes the value of the effective potential at the inner edge of the disk  
\begin{eqnarray}
	W-W_{in}+\int_0^P \frac{d P}{w}+\frac{\eta}{\eta-1}\frac{P_m}{w}=0.
	\label{EulerPot}
\end{eqnarray}

This equation represents the magneto-hydrostatic equilibrium, from which the equilibrium configurations of magnetized tori can be obtained  

\subsection{Construction of barotropic tori with constant angular momentum} 

On top of that, to evaluate the integral appearing in the expression (\ref{EulerPot}), two additional constraints to the model are required. Therefore, following the works of 
\cite{komissarov2006magnetized} and \cite{pimentel2018analytic}, we assume a torus where the fluid presents a constant angular momentum $l = l_0$, and which is described by a polytropic EoS for the pressure
\begin{eqnarray}
	P = Kw^\kappa
	\label{politrop},
\end{eqnarray} 
where $K$ is a constant and $\kappa$ is the polytropic index. In addition, \cite{komissarov2006magnetized} assumes a particular relationship for the magnetic pressure of the form 
\begin{eqnarray}
    P_m=K_m \mathcal{L}^{\eta - 1} w^\eta.
    \label{Pmfinal}
\end{eqnarray}
As a result, the Euler relativistic equation (\ref{EulerPot}) for a polytrope with constant angular momentum $l = l_0$ 
\begin{eqnarray}
	W-W_{in}+\frac{\kappa}{\kappa-1} \frac{P}{w}+\frac{\eta}{\eta-1} \frac{P_m}{w}=0.
	\label{EulerDef}
\end{eqnarray}

On the other hand, taking into account that   $u_t^2 = \mathcal{L}/\mathcal{A}$ where $\mathcal{A} = g_{t t} l_0^2+2 g_{t \phi} l_0+g_{\phi \phi}$, the effective potential can also be written as  
\begin{eqnarray}
	W (r,\theta)=\frac{1}{2} \ln |\mathcal{L} / \mathcal{A}|.
\end{eqnarray}

This way, we possess all the necessary ingredients to derive analytic solutions for magnetized, barotropic tori. Within this model, there are currently four free parameters: the potential at the inner edge of the torus, $W_{in}$, the angular momentum of the fluid, $l_0$, and the exponents of the expressions, $\kappa$ and $\eta$.

Nonetheless, it is necessary to determine the constants $K$ and $K_m$ that appear in the polytropic equation of state and the magnetic pressure expression (\ref{Pmfinal}). To this end, we define two additional parameters: the enthalpy at the disk center $w_c$ and the magnetization parameter at the disk center $\beta_c$, which allow us to determine the constants mentioned above. Consequently, it is necessary to specify the coordinate $r=r_c$ that defines the disk center. Following \cite{abramowicz1978relativistic}, we characterize the center of the disk as one of the two points in the equatorial plane where the fluid angular momentum $l_0$ equals the Keplerian specific angular momentum \cite{memmen2021geometrically}
\begin{eqnarray}
	l_K(r)^2=\left(1-\frac{2 M}{r}\right)^{-(1+2 q)} \frac{(1+q) M r^2}{r-(2+q) M}.
	\label{keplerianos}
\end{eqnarray}

Likewise, the other point where $l_0 = l_K$ is relevant because it defines the cusp of the disk, $r_{cusp}$. To differentiate which point corresponds to each feature, we use the criteria $r_{cusp} < r_c$ \cite{abramowicz1978relativistic}.  Regarding the other parameters, several conditions must be satisfied. For the fluid angular momentum $l_0$, we have that the disk is separated from the event horizon, or in the case of the $q$-metric, from  the singularity only if 
\begin{eqnarray}
	\left|l_0\right|>\left|l_{m s}\right|,
\end{eqnarray}
where $l_{ms}$ is the angular momentum corresponding to the marginally stable orbit   $r_{ms}$ \cite{abramowicz1978relativistic}. On top of that, we have that  
\begin{eqnarray}
	\left|l_0\right|>\left|l_{mb}\right|,
\end{eqnarray}
with $l_{mb}$, the angular momentum corresponding to the marginally bound orbit $r_{mb}$, for this condition, the disks are of finite extension only for the cases where  $W_{in} < 0$ \cite{komissarov2006magnetized}. However, these disks do not present a cusp on their inner edge  \cite{memmen2021geometrically}. Therefore, the disks of physical interest are the ones whose angular momentum satisfies the condition 
\begin{eqnarray}
	\left|l_{ms}\right|<\left|l_{0}\right| < \left|l_{mb}\right|,
\end{eqnarray}
due to them being separated from the singularity, presenting a centre and a cusp, and being of finite extension under the condition 
\begin{eqnarray}
	W_{c} < W_{in} < W_{cusp},
\end{eqnarray} 
where $W_{c}$ and $W_{cusp}$ are the values of the potential at the disk centre and on its cusp respectively. For the case of the $q$-metric, we obtain the values of $l$ from the expression of the specific angular momentum for a circular orbit (\ref{keplerianos}) evaluated at the radius corresponding to the marginally stable and bound orbits (Figure \ref{orbitscte}) \cite{memmen2021geometrically}. Then, the expression is evaluated as a function of the parameter $q$. The results obtained through this process are presented in Figure \ref{lorbitscte}.

Once the choice of the constant angular momentum of the fluid, $l_0$, is made, we can define the value of $W_{in}$ based on the criteria presented above. As a result, we have all the necessary elements to compute the physical variables that determine the disk configuration. To do this, we start from equation (\ref{EulerDef}) and use the definition of the magnetization parameter, $\beta = P/Pm$. Subsequently, we can find the values of gas pressure and magnetic pressure at the centre of the disk
\begin{align}
	&P_c  =w_c\left(W_{\text {in }}-W_c\right)\left(\frac{\kappa}{\kappa-1}+\frac{\eta}{\eta-1} \frac{1}{\beta_c}\right)^{-1}, 
	\label{presion}\\
	&P_{m_c}  =\frac{P_c}{\beta_c}.
	\label{presionm}
\end{align}

Using these values, the constants $K$ and $K_m$ are determined. Then, the enthalpy is determined as a function of the coordinates from equation (\ref{EulerDef}). Once the enthalpy is known, the magnetic pressure is computed using  (\ref{Pmfinal}) and the gas pressure with (\ref{politrop}). Finally, the rest mass density is obtained through 
\begin{eqnarray}
    \rho=w-\frac{\kappa \mathrm{P}}{\kappa-1},
	\label{masarep}
\end{eqnarray}
with the equation (\ref{masarep}), we can describe $\rho$ as a function of the coordinates $\rho(r,\theta)$, which allows us to show its behaviour in a plane.  Moreover, it is noteworthy to characterize the shadow of such configurations through numerical simulations.  

\section{Magnetized Disks} \label{ResI}

To investigate magnetized geometrically thick disks surrounding a naked singularity described by the $q$-metric, we revisit the work conducted by \cite{faraji2021magnetised} and systematically construct a series of equilibrium configurations as examples of these tori with varying angular momentum parameters. This exploration aims to examine how the mass quadrupole, a characteristic of this spacetime, and other model parameters influence the disk morphology. Our analysis presents the particularity of considerating the role of the multipole moments (\ref{multipole}) and their physical interpretation in the disk construction. Therefore, we fix the value of the mass monopole $M_{0}$, which coincides with the ADM mass of the spacetime \cite{boshkayev2016motion}. This approach allows us to assess the impact of the parameter $q$ on a spacetime where the compact object possesses a specific mass, facilitating comparisons with astronomical observations.

Now, with the aim of studying the effect of the quadrupole on the accretion disks, we chose a series of configurations that allow us to explore the wide spectrum of possibilities related to the deviation of the object from spherical symmetry. For this study, we chose two values of $q = − 0.3$ and $q= − 0.15$, which correspond to objects with prolate deformation, following \eqref{multipole}, whereas, for the oblate case, we take the value $q = 0.1$. Besides, we use the case $q=0$ as a reference, as it reproduces Schwarzschild's spacetime, for which geometrically thick disks endowed with a magnetic field have been extensively studied \cite{komissarov}.  

Likewise, three cases of the magnetization parameter at the disk's center ($\beta_c =$ {1, 0.1, 0.01}) are presented for the selected set of $q$ values. This allows for the examination of the impact of the magnetic field on disk formation in this spacetime. According to the convention adopted to define this parameter, the first case represents a disk where the magnetic pressure, $P_m$, equals the hydrostatic pressure, $P$. On the other hand, the other two cases correspond to tori dominated by magnetic effects, with the effect being more pronounced in the last case.

\begin{figure*}
	\centering
	\begin{subfigure}[b]{0.47\textwidth}
		\includegraphics[width=\textwidth]{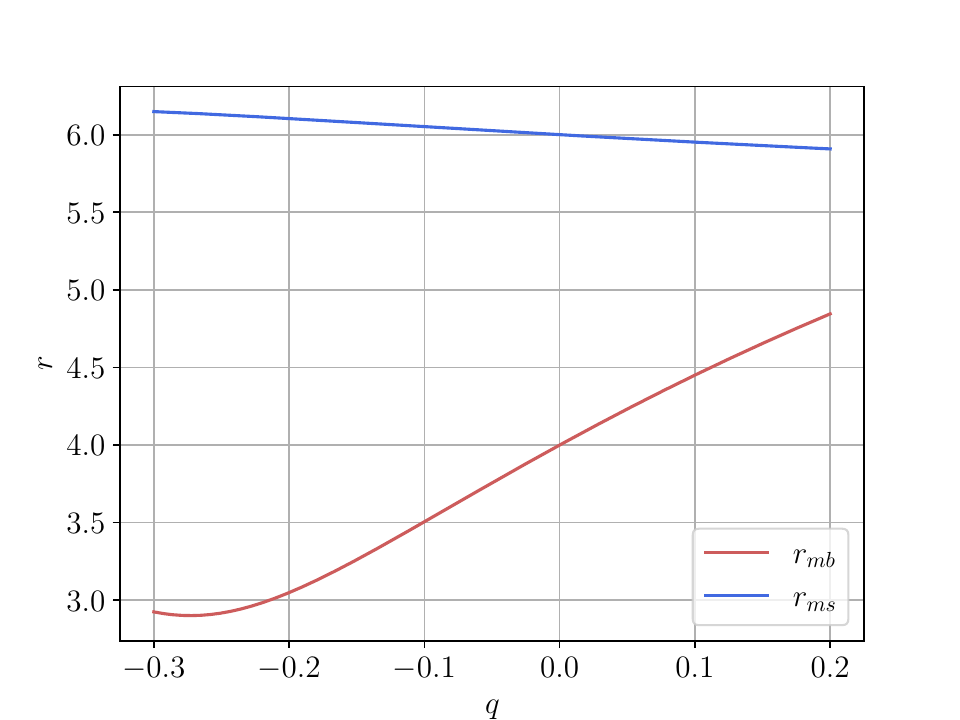}
		\caption{}
		\label{orbitscte}
	\end{subfigure}
	\begin{subfigure}[b]{0.47\textwidth}
		\includegraphics[width=\textwidth]{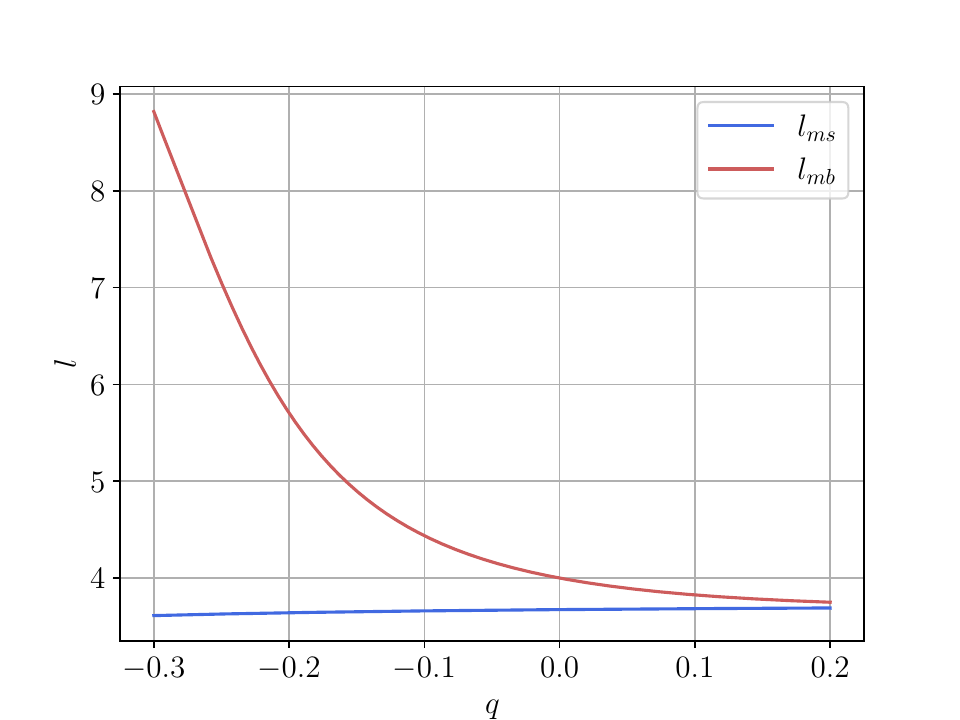}
		\caption{}
		\label{lorbitscte}
	\end{subfigure}
	\caption{(a) Marginally stable and marginally bound orbits as a function of the parameter $q$. (b) Specific angular momentum of the marginally stable and marginally bound orbits as a function of the parameter $q$.}
\end{figure*}

To select the values of the additional parameters for the disk construction, we refer to the analysis carried out in section \ref{construc}. For the $q$-metric, the behaviour of the orbits of interest is depicted in Figure \ref{orbitscte}, while Figure \ref{lorbitscte} presents the angular momenta. From Figure \ref{orbitscte}, we observe that the marginally stable orbit exhibits a slightly decreasing tendency, whereas the marginally bound presents an increasing behaviour. Therefore, the distance between the two interest radii reduces when the value of $q$ increases. It is worth emphasizing that the marginally bound orbit $r_{mb}$ is defined only for values of $q > -0.3$. Hence, we select this threshold as the limit for constructing tori around a naked singularity described by the $q$-metric.

On the other hand, from Figure \ref{lorbitscte}, we observe that the angular momentum of the stable orbit $l_{ms}$, has a nearly constant behavior, while for the bound orbit $l_{mb}$ it rises rapidly when the value of $q$ decreases. Besides, we also observe that for $q>0.1$, the values of the angular momenta almost even out,  imposing strong restrictions on disk construction within this range. This distinct behavior of the previous orbits and angular momenta, compared to precedent studies on thick disks in this spacetime \cite{memmen2021geometrically, faraji2021magnetised}, can be attributed to the fixed ADM mass of the compact object, which imposes additional constraints on the value of the deformation parameter $q$. Table \ref{tablaADMcte} displays the employed parameters for the disk construction. On top of that, it is relevant to remark that the parameters $\kappa = \eta = 4/3$ y $w_c = 1$ are used for each one of the configurations. 
\begin{table}
	\centering
	\caption{Paramaters for tori construction}
	\begin{tabular}{ccccccc}
		\hline $q$ & $l_0$ & $r_{\text {cusp }}$ & $r_{\mathrm{c}}$ & $W_{\text {cusp }}$ & $W_{\mathrm{c}}$ & $W_{\text {in }}$ \\
		\hline  $-0.3$ & $4.5$ & $3.64$ & $15.832$ & $0.011$ & $-0.031$ & $-0.025$  \\[1mm]
		$-0.15$ & $4.20$ & $3.78$ & $12.64$ & $0.075$ & $-0.037$ & $-0.030$ \\[1mm]
		$0$ & $3.95$ & $4.11$ & $9.97$ & $-0.012$ & $-0.045$ & $-0.030$ \\[1mm]
		$0.1$ & $3.82$ & $4.47$ & $8.44$ & $-0.040$ & $-0.050$ & $-0.048$ \\
		\hline
	\end{tabular}
	\label{tablaADMcte}
\end{table}

\begin{figure*}
	\centering
	\begin{subfigure}[b]{0.47\textwidth}
		\includegraphics[width=\textwidth]{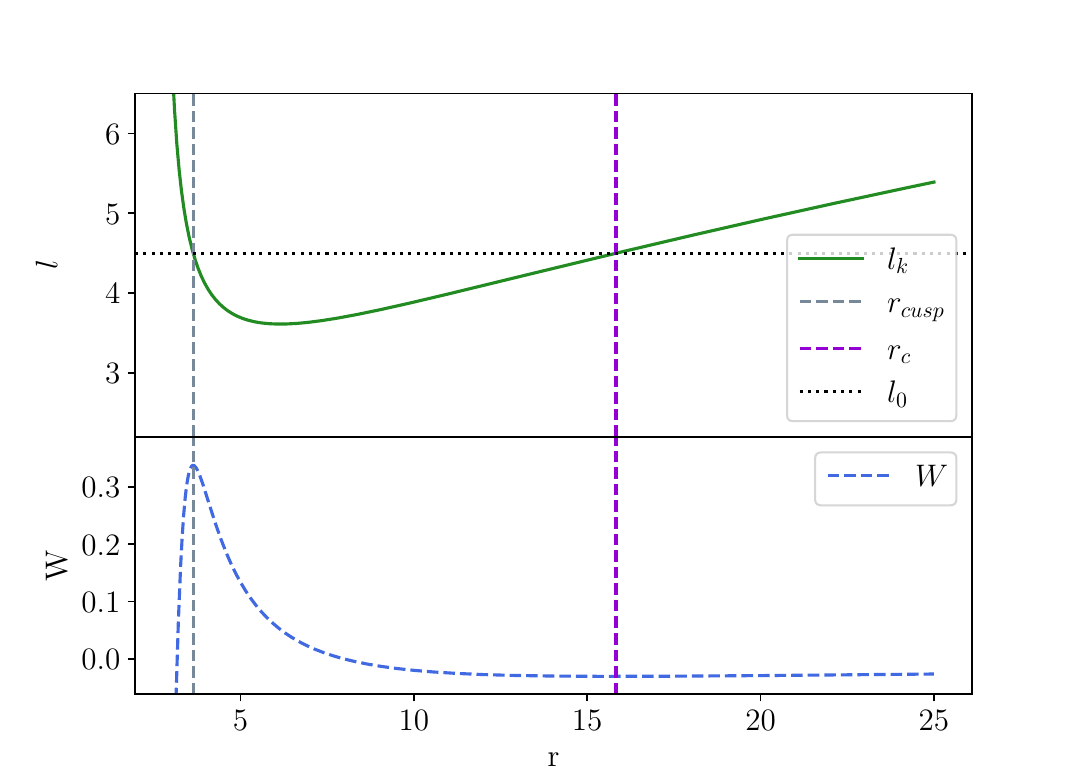}
		\caption{}
		\label{fig:vsq03}	
	\end{subfigure}
	\begin{subfigure}[b]{0.47\textwidth}
		\centering
		\includegraphics[width=\textwidth]{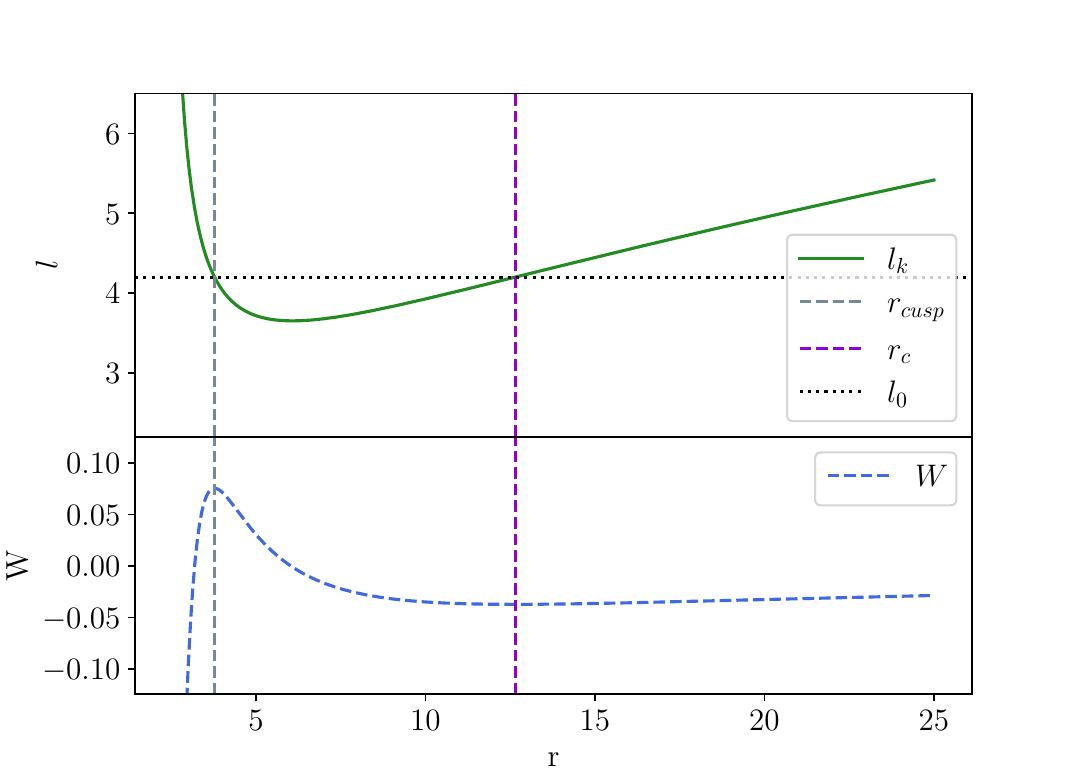}
		\caption{}
		\label{fig:vsq015}
	\end{subfigure}
	\begin{subfigure}[b]{0.47\textwidth}
		\centering
		\includegraphics[width=\textwidth]{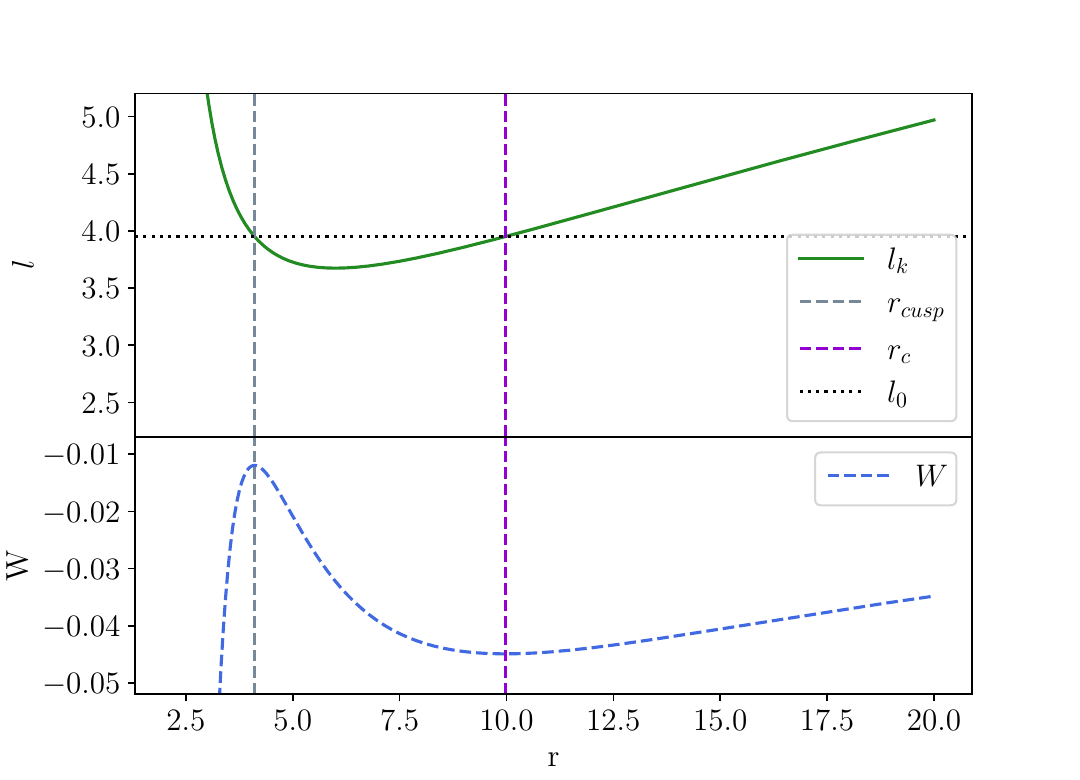}
		\caption{}
		\label{fig:vsq0}
	\end{subfigure}
	\begin{subfigure}[b]{0.47\textwidth}
		\includegraphics[width=\textwidth]{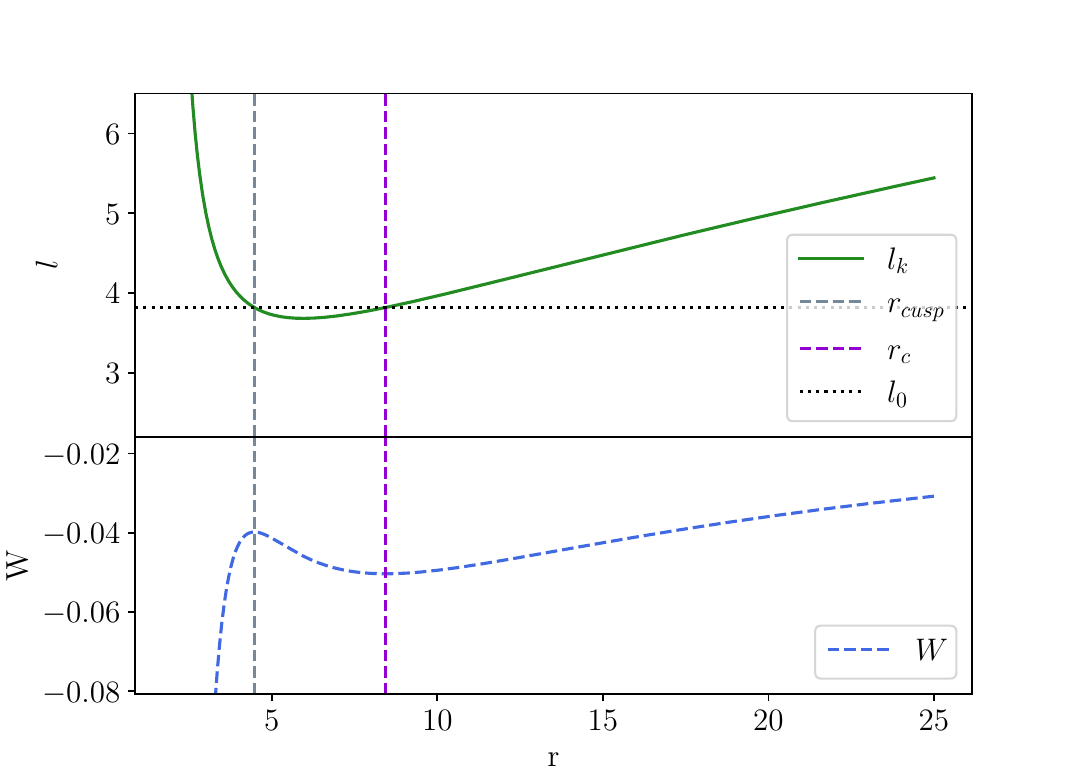}
		\caption{}
		\label{fig:vsq01}	
	\end{subfigure}
	\caption{Keplerian angular momentum and effective potential as a function of $r$. (a) $q = -0.3$. (b) $q = -0.15$. (c) $q = 0$. (d) $q = 0.1$.}
	\label{fig:lkepandW}
\end{figure*}

Figure \ref{fig:lkepandW} illustrates the behaviour of the Keplerian angular momentum and the effective potential in the equatorial plane ($\theta = \pi/2$) for the different elections of $q$. From these, we see that the centre and the cusp of the disk are determined as the points where the angular momentum of the fluid $l_0$ matches the Keplerian angular momentum,  which coincide respectively with a minimum and maximum of the effective potential $W$. Notably, the behaviour of these variables is not affected by the magnetization parameter $\beta_c$, and the graphs remain the same for a fixed selection of $q$ and $l_0$.

In greater detail, figure \ref{fig:vsq03} shows a rapidly decaying potential $W$ and a considerable separation between the cusp and the centre of the disk, which indicates a highly extended configuration.  In contrast, reviewing figures \ref{fig:vsq015}, \ref{fig:vsq0}, and \ref{fig:vsq015} indicates a trend toward decreasing distance between these two points, resulting in more compact disks as $q$ increases. Nonetheless, it is important to note that this effect cannot be solely attributed to the deformation parameter, as it may also be associated with the decrease in the permitted values of $l_0$ when $q$ increases, as demonstrated in Table \ref{tablaADMcte}. 

\begin{figure*}
	\centering
	\includegraphics[width=\textwidth]{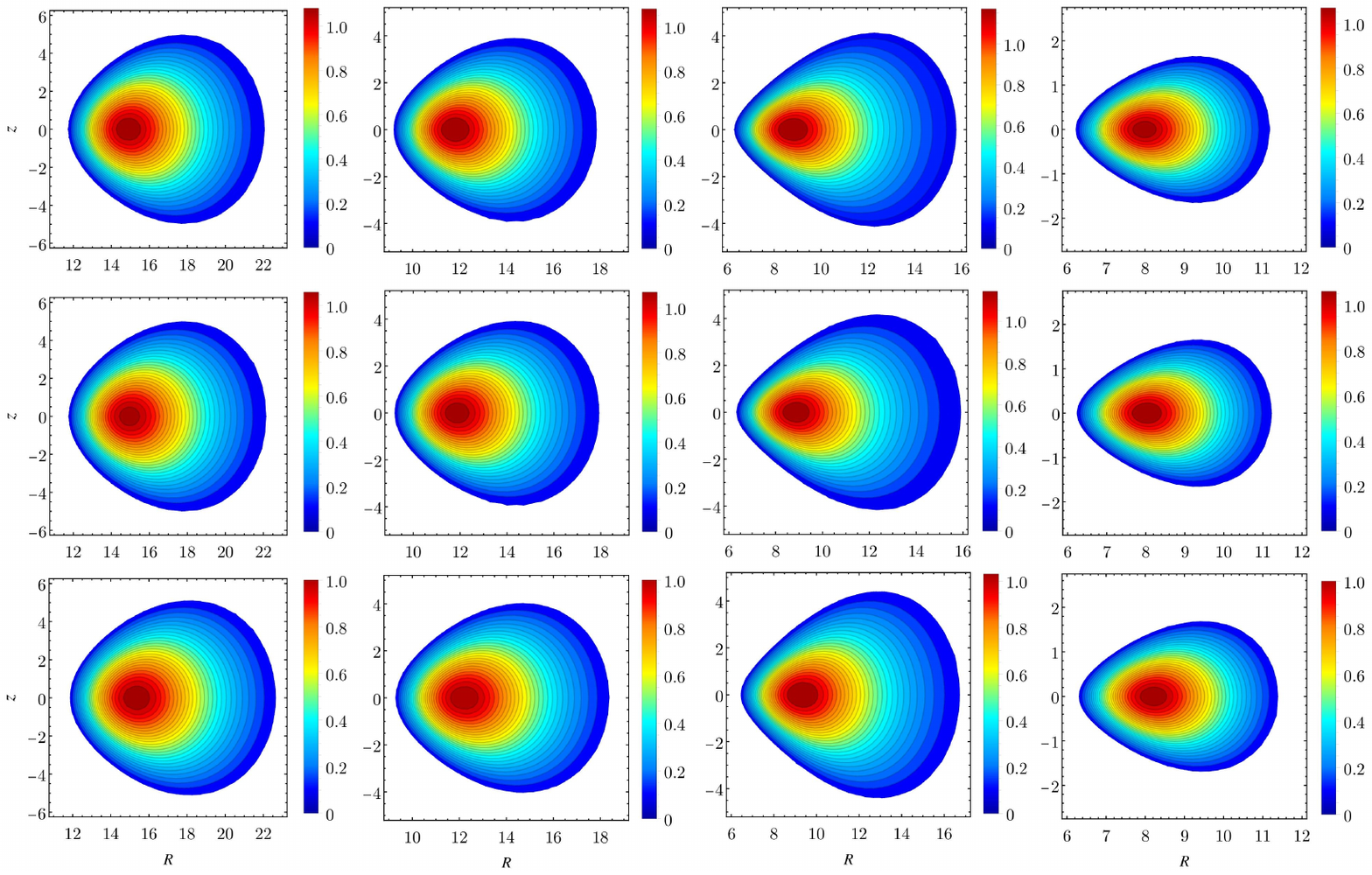}
	\caption{Mass density isocontours. Each column corresponds to a different value of $q =\{-0.3, -0.15, 0, 0.1\}$ (From left to right). Rows correspond to values of the magnetization parameter $\beta_c=$ $\{0.01, 0.1, 1\}$ (From top to bottom).}
	\label{Fig:Isocontours}
\end{figure*}

On the other hand, Figure \ref{Fig:Isocontours} showcases a panel displaying isocontours of mass density in the plane for our different selections of $q$ and $\beta_c$. Each row in the image corresponds to a different value of $\beta_c = \{0.01, 0.1, 1\}$ from top to bottom, while each column represents a value of $q = \{-0.3, -0.15, 0, 0.1\}$ from left to right. In the case of $\beta_c = 1$ and the four corresponding $q$ values, corresponding to the entire last row in Figure \ref{Fig:Isocontours}, we observe that the disk with the greatest extent is associated with $q = -0.3$, located on the left-most side. Moreover, there is a discernible trend of increasing compactness as the value of $q$ increases (This is accompanied by a decrease in the magnitude of the angular momentum). Furthermore, the prolate or oblate character of the compact object deformation has an impact on the disk's shape. For $q < 0$ (prolate deformation), the disks exhibit a more rounded shape, while for $q > 0$ (oblate deformation), the disks resemble the shape of a seed.

In general, we find that the disks are more extended and farther away from the compact object as the value of $q$ decreases and the angular momentum $l_0$ increases (Going from the right to left); we cannot individualize the impact of each parameter since each configuration presents a different combination of $q$ and $l_0$. As noted in Table \ref{tablaADMcte}, $l_0$ has the highest value for $q=-0.3$ and decreases as $q$ increases. Although, we encounter one shared result with \cite{faraji2021magnetised}, since for prolate configurations we find that the disk is enlarged in the vertical direction near the inner edge, whereas, for the oblate case, the disk tends to be flattened out around the equator. Now, regarding the comparison with $q=0$ (third disk, from left to right), the disk presented for $q =-0.3$ does not surpass the extension in the radial direction $R$, of the control case disk $q=0$ by a wide margin; while for $q=-0.15$, a smaller disk extension is obtained.

Next, examining the first and second row in Figure \ref{Fig:Isocontours}, we can identify the disks formed for values of $\beta_c = 0.01 $ and $\beta_c = 0.1$, respectively. Upon comparison, we can observe the previously mentioned effects of the $q$ parameter and the angular momentum $l_0$ on the disks. Besides, we note that a decrease in $\beta_c$ leads to an increase in the maximum value of density; however, this phenomenon is quite mild. We found that the maximum density is higher for magnetically dominated disks. However, it is observed that the most significant variation occurs between the first two cases ($\beta_c = 1$ and $\beta_c = 0.1$), showing that increasing the degree of magnetization once it is already dominant does not have a significant effect on the mass density. Likewise, the shifting of said maximum towards the cusp of the disk is observed,  indicating a greater concentration of mass in the interior of the disk as magnetic dominance increases \cite{faraji2021magnetised, komissarov2006magnetized}. 

Furthermore, when observing the graphs of the disks as $\beta_c$ varies, i.e. comparing the different rows in Figure \ref{Fig:Isocontours}, it can be seen that the approaching trend of the disks to the compact object as the value of $q$ increases and $l_0$ decreases, is maintained; while analyzing configurations such as those of the extremum values of $\beta_c$ for $q= -0.3$ (Top and bottom of the first column) we observe that the extension of the tori is very similar. This analysis confirms that the magnetization parameter primarily influences the matter distribution within the disk rather than its specific geometric structure \cite{faraji2021magnetised}.

On the other hand, we can also observe that the maximum values of the density are always presented for Schwarzschild's spacetime ($q=0$), whereas for spacetimes with non-vanishing quadrupole ($q \neq 0$) the peak value is not as high. Now, regarding the relationship between the extension of the disk and its mass density top value, we see that even though tori with a combination of negative values of $q$ and a higher angular momentum $l_0$ are considerably more extensive than those with positive values of $q$ and correspondingly lower values of $l_0$, the maximum values do not variate significantly. Thus, this property is said to be related to a great extent to the mass of the compact object itself and not to the extension of the torus.

\section{Radiative Transfer Equations}
\label{sec:RTE}

After constructing a series of equilibrium configurations for magnetized tori in the $q$-metric spacetime, our focus shifted to assessing their potential impact on observations. To achieve this, we studied the behaviour of photons that successfully escape the gravitational pull of the compact object (In this case, a naked singularity). This analysis enabled us to generate an image of the surrounding accretion disk and measure its emission spectra.  By doing so, we aimed to provide an initial characterization of observable parameters that can be detected by telescopes here on Earth.

Considering this, it is relevant to present a brief overview of the model used to describe radiative transfer, which follows the procedure 
presented by \cite{velasquez2023synchrotron}. In this case, the radiative transfer is described using the covariant formulation \cite{rybicki}
\begin{equation}
    \frac{d}{d\lambda}\left( \frac{I_\nu}{\nu^3} \right) = \left( \frac{j_\nu}{\nu^2} \right) - (\nu\alpha_\nu)\left( \frac{I_\nu}{\nu^3}\right), \label{eq:radtrans}
\end{equation}
where $I_{\nu}$, $j_{\nu}$, and $\alpha_{\nu}$ represent the specific intensity, emission coefficient, and absorption coefficient, respectively, which are measured in the comoving frame, $\lambda$ is an affine parameter and the terms in parentheses are Lorentz invariants. Likewise, $\nu$ is the frequency, $\lambda$ is an affine parameter, and the quantities in parentheses are Lorentz invariants. Moreover, it is convenient to introduce the definition of the optical depth $\tau_\nu$
\begin{equation}
    \tau_\nu  = \int_{\lambda_0}^{\lambda}\nu\alpha_\nu d\lambda', \label{eq:tau}
\end{equation}
this allows us to classify the tori into two categories: optically thick if $\tau_\nu$ = 0 and optically thin if $\tau_\nu$ > 0. Upon introducing this definition and defining the source function as $S_\nu = j_\nu/\alpha_\nu$, which does not depend on  $\tau_\nu$, equation \ref{eq:radtrans} can be rewritten as
\begin{equation}
     I_\nu(\tau_\nu) = I_\nu(0)\text{e}^{-\tau_\nu} + S_\nu(1 - \text{e}^{-\tau_\nu}). \label{eq:intensity}
\end{equation}

In general, emission and absorption coefficients depend on the radiative process. For our study, we opted to examine the effect of non-thermal synchrotron radiation, which can be described as a power law with constant coefficients
\begin{equation}
    j_\nu \propto B^{(\gamma+1)/2}\nu^{(1-\gamma)/2}, 
    \quad
    \alpha_\nu \propto B^{(\gamma+2)/2}\nu^{-(\gamma+4)/2}, \label{eq:coefficents}
\end{equation}
here, $\gamma = 2s + 1$, where $s = 0.75$ is the spectral index, whose value is motivated by observations of S-type radio sources \cite{pacholczyk1970non}. Then, through Lorentz invariants, we can determine the specific intensity measured by a distant observer
\begin{equation}
    I_{\nu_{\text{obs}}} = \left( \frac{\nu_{\text{obs}}}{\nu_{\text{em}}} \right)^3 I_{\nu_{\text{em}}}  = g^3 I_{\nu_{\text{em}}}, \label{eq:iobs}
\end{equation}
in this expression, the subscripts "obs" and "em" refer to the intensity and frequency received by the observer and emitted by the disk, respectively. Additionally, the term $g = (1+z)^{-1}$  is known as the red-shift factor, which can be calculated as
\begin{equation}
    g = \frac{\nu_{\text{obs}}}{\nu_{\text{em}}} = \frac{-p_{\mu}v^\mu|_{\lambda_\text{obs}}}{-p_{\mu}u^\mu|_\lambda}.
\end{equation}

In the latter, $\left\lbrace v^\mu \right\rbrace = \left\lbrace-1,0,0,0\right\rbrace$ is the 4-velocity of the distant observer, while $p_\mu$ are the components of four-momentum for photons. 

\section{Numerical Setup}
\label{sec:NS}

We carried out numerical simulations using \texttt{OSIRIS} ({\bf O}rbits and {\bf S}hadows {\bf I}n {\bf R}elativ{\bf I}stic {\bf S}pace-times) \cite{velasquez2022osiris}, a code of our authorship based on the backward ray-tracing algorithm for stationary and axially-symmetric space-times. \texttt{OSIRIS} evolves null geodesics “backwards in time” by solving the equations of motion in the Hamiltonian formalism. Our code works based on the image-plane model, an assumption where photons came from the observation screen in the direction of the compact object. Some authors employ this method to simulate shadows around black holes \cite{Johannsen_2013, doi:10.1142/S0218271816410212, disformal}, classifying the orbits of photons into two groups: those that reach the event horizon and those that escape to infinity. Every pixel on the image plane corresponds to an initial condition for each photon. Thus, when the code assigns a colour to each pixel according to the initial conditions, we obtain an intensity map corresponding to the image of the radiation coming from the accretion disk around the compact object. 

The null-geodesic integration process is performed for two different angles measured from the pole, $\theta_0 = 85^\circ$ and $\theta_0 = 30^\circ$, the latter, motivated by the EHT observations of the compact object Sgr A* \cite{akiyama2022first}. For this process we assume that the Minkowskian observer is located at $\left\lbrace t_0, r_0, \theta_0, \phi_0 \right\rbrace $ = $\left\lbrace 0, 1000, \theta_0, 0 \right\rbrace$. We define the surface of the tori, as the two-dimensional region where $\rho = 0.01$. Once the photon reaches this surface, the integration process of the radiative equation starts and keeps going while the photon remains inside the tori. For an optically thick disk $\tau_\nu$ = 0, implying that $I_{\nu_{\text{obs}}} \propto g^3$. For an optically thin disk, the algorithm calculates the value of $\lambda_0$ once the photon reaches the disk surface. Then, we use an Euler method to solve equation \eqref{eq:tau} along the photon path inside the disk. Finally, the observation screen has a range of $-12 \le x, y \le 12$ with a resolution of $1024\times 1024$ pixels for all simulations of the intensity map of the torus.


\section{Torus spectra} \label{ResII}

\subsection{Emission profile}

The emission profiles were obtained by computing the flux from the tori, whether optically thick or thin. In order to do this, we used the expression for the flux 
\begin{equation}
	F_\nu=\int I_\nu \mathrm{~d} \Omega.
	\label{flujo}
\end{equation}

\begin{figure*}
	\centering
	\includegraphics[width=\textwidth]{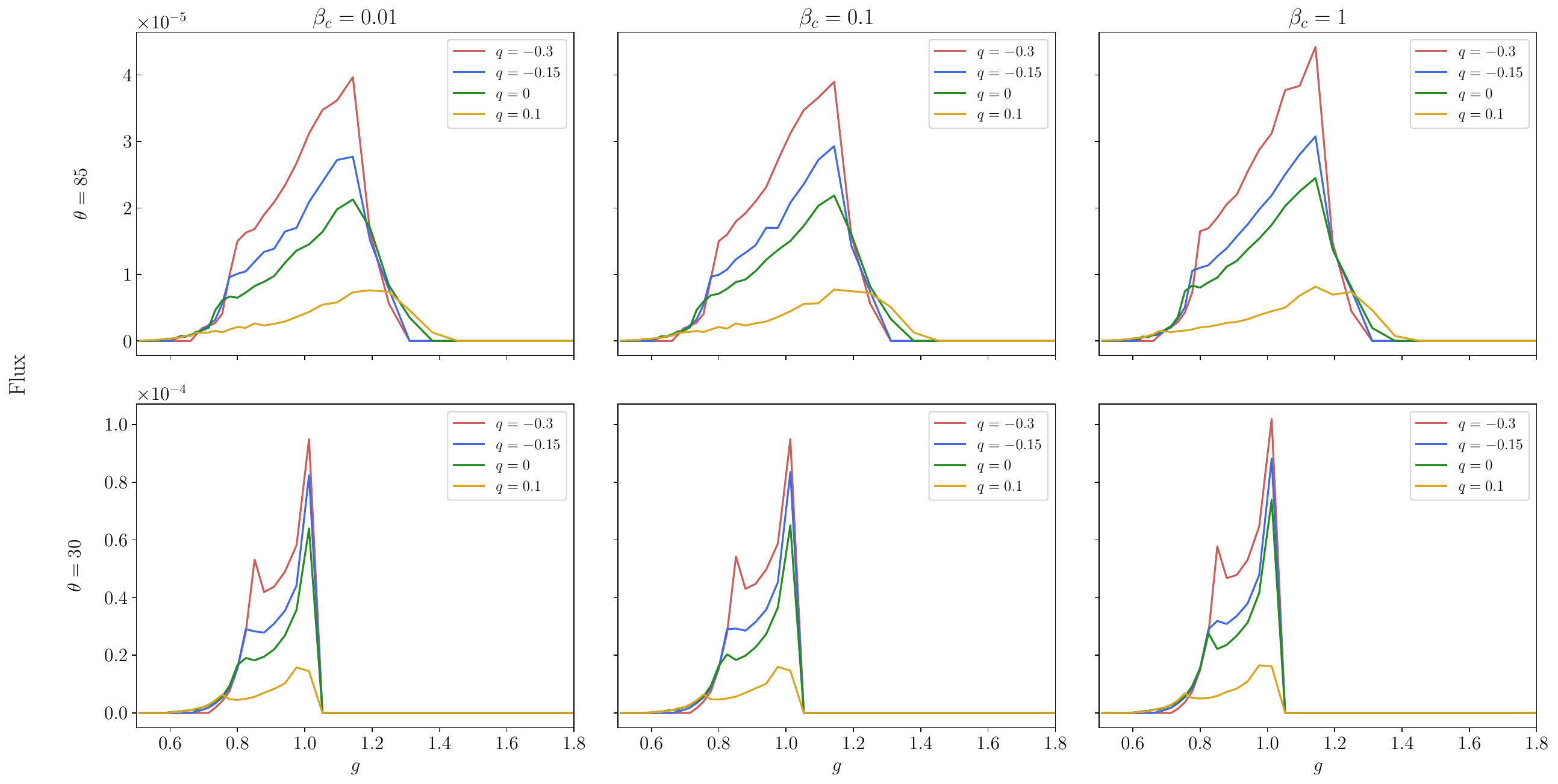}
	\caption{Spectra of optically thick disks for values of $q = \{-0.3, -0.15, 0, 0.1 \}$ and with magnetization parameter $\beta_c = \{1, 0.1, 0.01\}$. The rows correspond to values of the observation angle $\theta=$ 85° (Top) and $\theta=$ 30° (Bottom).}
	\label{EspThick}		
\end{figure*}

\begin{figure*}
	\centering
	\includegraphics[width=\textwidth]{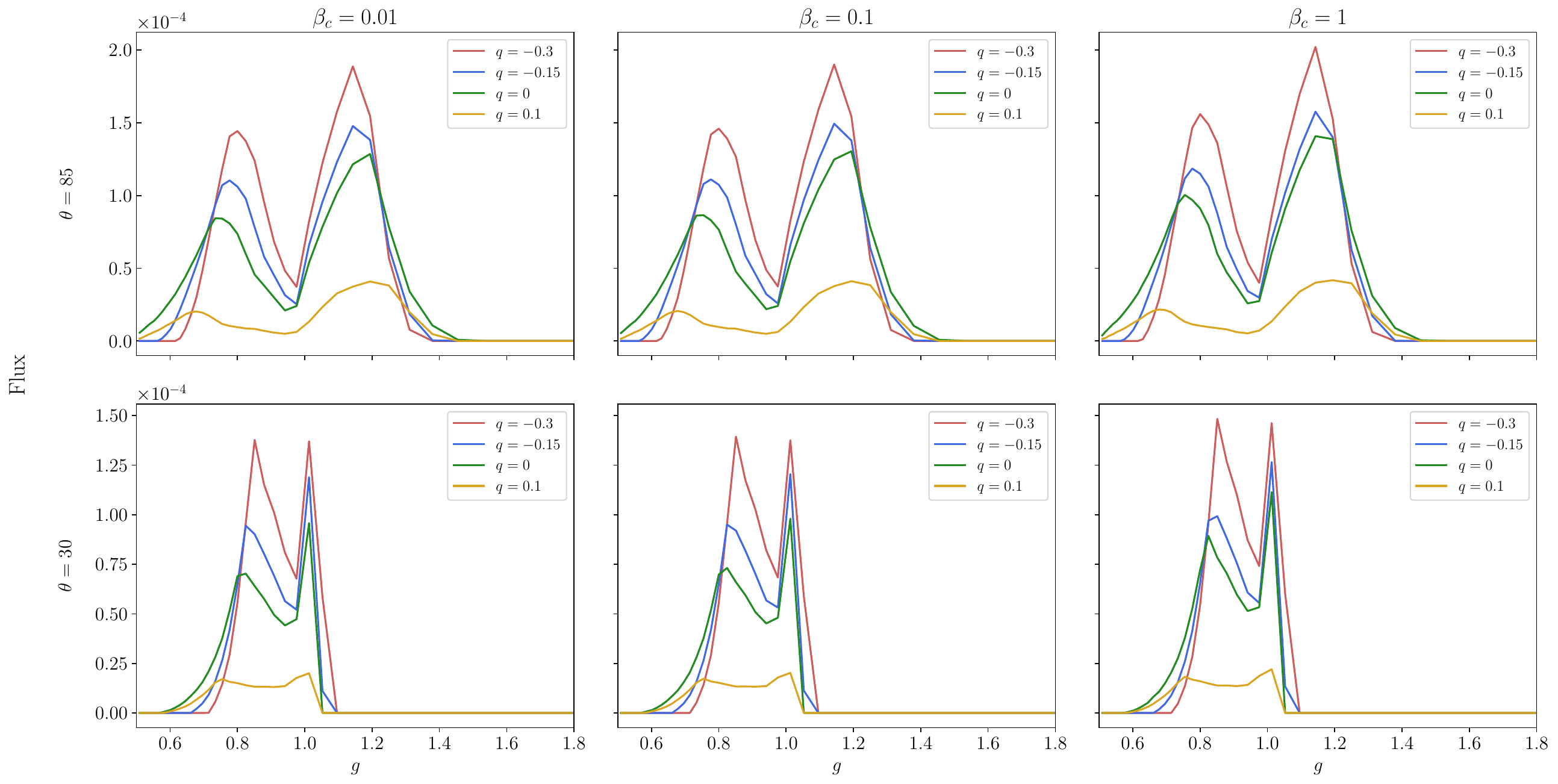}
	\caption{Spectra of optically thin disks for values of $q = \{-0.3, -0.15, 0, 0.1 \}$ and with magnetization parameter $\beta_c = \{1, 0.1, 0.01\}$. The rows correspond to values of the observation angle $\theta=$ 85° (Top) and $\theta=$ 30° (Bottom).}
	\label{EspThin}		
\end{figure*}

This expression is valid for both categories of optical depth, where $d\Omega$ represents the solid angle element. Since $\texttt{OSIRIS}$ uses an image-plane model, we can express this quantity as $d\Omega = dx dy / r_0^2$, which is fully characterized by defining the simulation resolution. Furthermore, to calculate the observed flux computationally, we established a grid of frequencies ranging from $\nu_i = 0.5$ to $\nu_f = 2.0$ with a resolution of $d\nu = 0.0375$. This allows the flux computation to be reduced to a sum of the individual contributions of each frequency to the intensity. The criterion used to determine the frequency at which a photon is emitted at each point in space is also defined as follows
\begin{equation}
	\left|\nu_{\text {grid }}-\nu_{\text {em }}\right| < \delta_\nu ,
\end{equation}
where  $\delta_\nu = 0.2$ is the tolerance value selected for this study. In Figures \ref{EspThick} and \ref{EspThin}, the emitted fluxes for optically thick and thin disks are portrayed, respectively, in the cases of  $q = \{$-0.3, -0.15, 0, 0.1$\}$ for values of $\beta_c = \{$1, 0.1, 0.01$\}$ and observation angles of  $\theta = \{$85°, 30°$\}$, as a function of the redshift factor $g$. From these graphs, we note that the maximum flux values occur for the case $q=-0.3$, which is the disk with the largest extension due to a combination of a negative value of $q$ and a higher angular momentum $l_0$, as reported previously. Following closely in flux magnitude is the case of $q=-0.15$. Despite having a slightly smaller extension than a torus formed in Schwarzschild spacetime, it achieves a greater flux magnitude.

These findings suggest a correlation between the observed flux and the compound influence of the value of the parameters $q$ and $l_0$ over the disk extension. However, in the case of a disk with $q=0.1$, the spectrum generated has a significantly smaller magnitude compared to the other cases. This result is associated with the reduced extension of this torus in comparison with its counterparts, as shown in Figure \ref{Fig:Isocontours}. Notably, this behaviour of the emission profiles with respect to the parameters $q$ and $l_0$ is consistent across all the values of $\beta_c$ examined, as well as with the proposed observation angles $\theta$.

On the other hand, upon examining the impact of the magnetization, we found that the maximum intensity in both peaks increases with the magnetization parameter across all cases.  This effect is particularly noticeable when transitioning from magnetically dominated disks ($\beta_c = 0.1$) to disks where magnetic and hydrodynamic pressure are of the same order ($\beta_c = 1$). Said behaviour might be attributed to the chosen model for synchrotron radiation. The coefficients in equation (\ref{eq:coefficents}) show that the magnetic field magnitude depends, on its power, on the spectral index $s$. However, the source function $S$, defined as the ratio between the emission and absorption coefficients, does not depend on this index. Instead, we found that the magnetic field always decreases as $B^{-1/2}$. Therefore, as the magnetic field increases, the contribution of the source function to the intensity decreases, as evidenced in (\ref{eq:intensity}). Given this and the fact that decreasing $\beta_c$ increases the magnetic pressure, and hence the magnetic field, it is consistent that the flux decreases when $\beta_c$ becomes smaller.

Moving onto optically thick disks, we can see in Figure \ref{EspThick} that for $q=0$, the shape of the emission profile matches those found by \cite{pariev1998line, wu2007iron} for thick disks, with a slight peak on the left for $\theta =$ 85°, which becomes more pronounced as the observation angle decreases. In addition, in this graph, we notice that the flux reaches higher values when $\theta =$ 30°,  this is because near the equatorial plane ($\theta =$ 85°), the observable region of the disk is smaller. This phenomenon is known as self-eclipsing and is visible in Figure \ref{thick85}.

Optically thick disks have a distinct emission profile with two peaks. The first peak corresponds to redshifted emission from the part of the disk moving away from the observer, while the second peak on the right corresponds to blueshifted emission from the material moving towards the observer \cite{fabian1989x}. Comparing the two rows in Figure \ref{EspThick}, we can see that the peak corresponding to redshifted emission is almost imperceptible for $\theta=85^\circ$, making the profile appear to have a single peak. This behaviour is explained by the self-eclipsing mentioned earlier.       

On the other hand, when reviewing the behaviour of the emission for optically thin disks in Figure \ref{EspThin} we find that the profile is characterized by two peaks corresponding again, to redshifted and blueshifted emission, respectively, as found for opaque disks. Nevertheless, in this case, the maximum associated with the redshifted emission is much more noticeable. This is due to the very nature of this kind of tori, which makes the self-eclipsing not as relevant. This is consistent with the previous description of translucent disks, for which the photons travel an optical path within the disk before exiting it, resulting in an increased contribution to the flux.

Additionally, another consequence of this nature is that the magnitude of the spectrum is similar for both observation angles, unlike thick disks where self-eclipsing generated a considerable difference between the magnitude of the profiles. Moreover, due to a larger portion of the torus being observable, the order of magnitude of the emission is higher in thin disks, either for $\theta = 30$° or $\theta = 85$°.

Furthermore, when examining the observation at 30° in Figure \ref{EspThin}, a striking feature becomes apparent for tori in this spacetime. Specifically, for the combination of $q = -0.3$ and a higher angular momentum $l_0$, resulting in a larger extension, the maximum associated with redshift surpasses the magnitude of the blueshift. This differential phenomenon is noteworthy when comparing it with other configurations, such as the one obtained in this study for Schwarzschild's spacetime. Last, it is worth noting that for all cases, changing the observation angle from 85° to 30° results in a blueshift in the peak location, which has been previously reported in the literature \cite{jovanovic2012broad}.
 
\subsection{Intensity maps}

\begin{figure*}
	\centering
	\begin{subfigure}[b]{0.48\textwidth}
		\includegraphics[width=\textwidth]{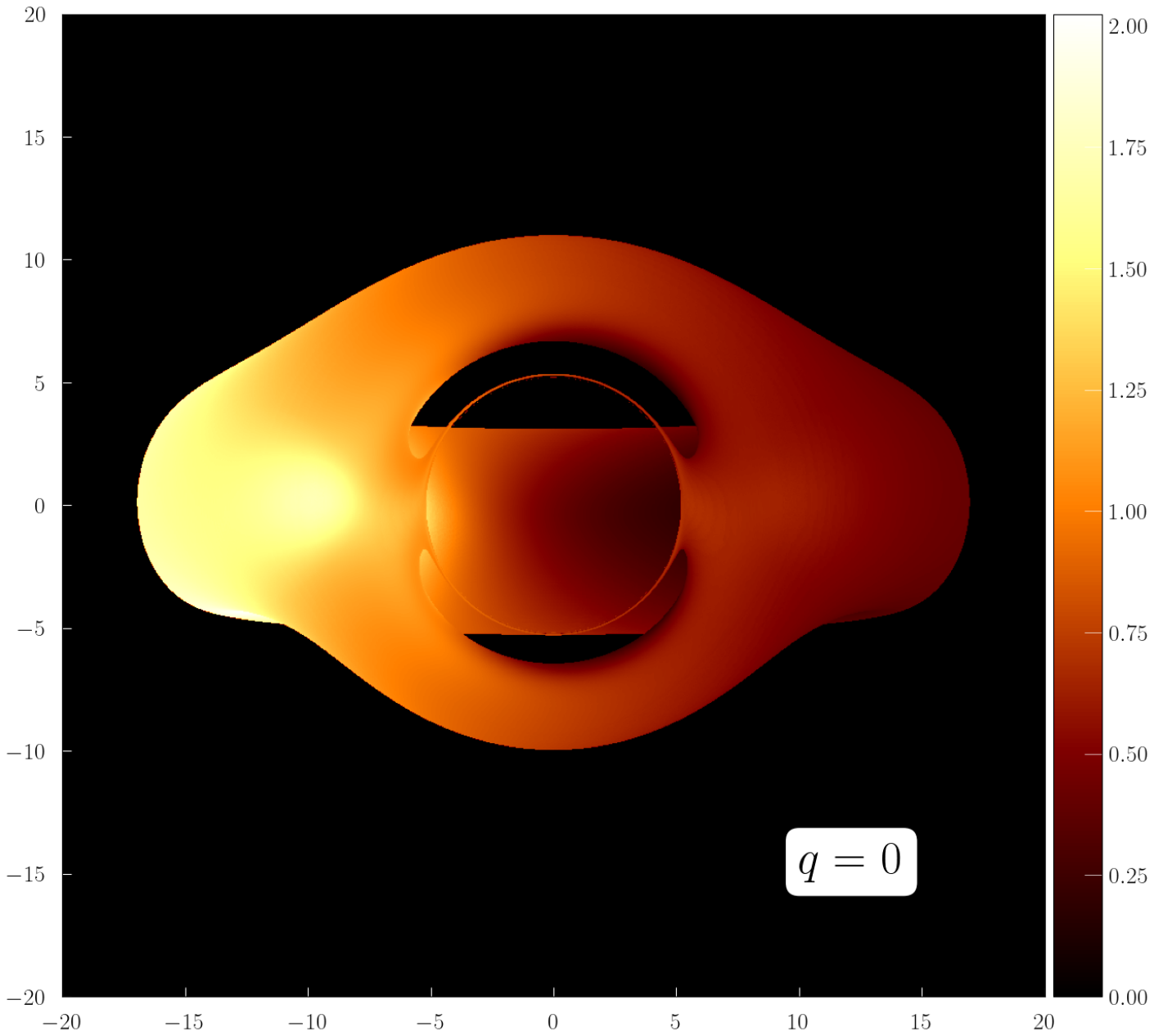}
		\caption{}
		\label{thin85q0}
	\end{subfigure}
	\begin{subfigure}[b]{0.48\textwidth}
		\includegraphics[width=\textwidth]{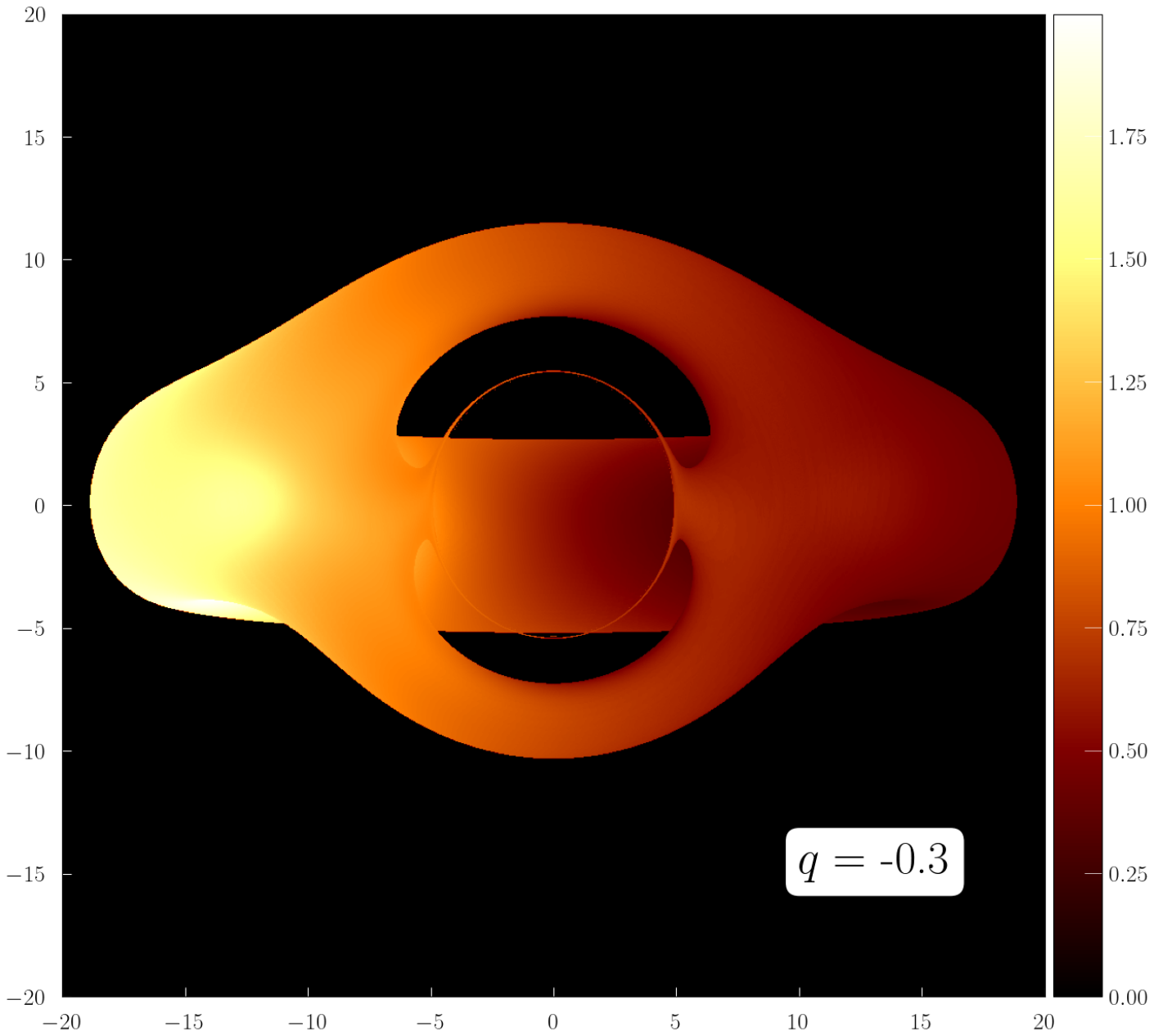}
		\caption{}
		\label{thin85q03}
	\end{subfigure}
	\caption{Optically thin magnetized tori around a naked singularity seen from $\theta=85$°. (a) $q=0$. (b) $q=-0.3$.}
	\label{thin85}
\end{figure*}

\begin{figure*}
	\centering
	\begin{subfigure}[b]{0.48\textwidth}
		\includegraphics[width=\textwidth]{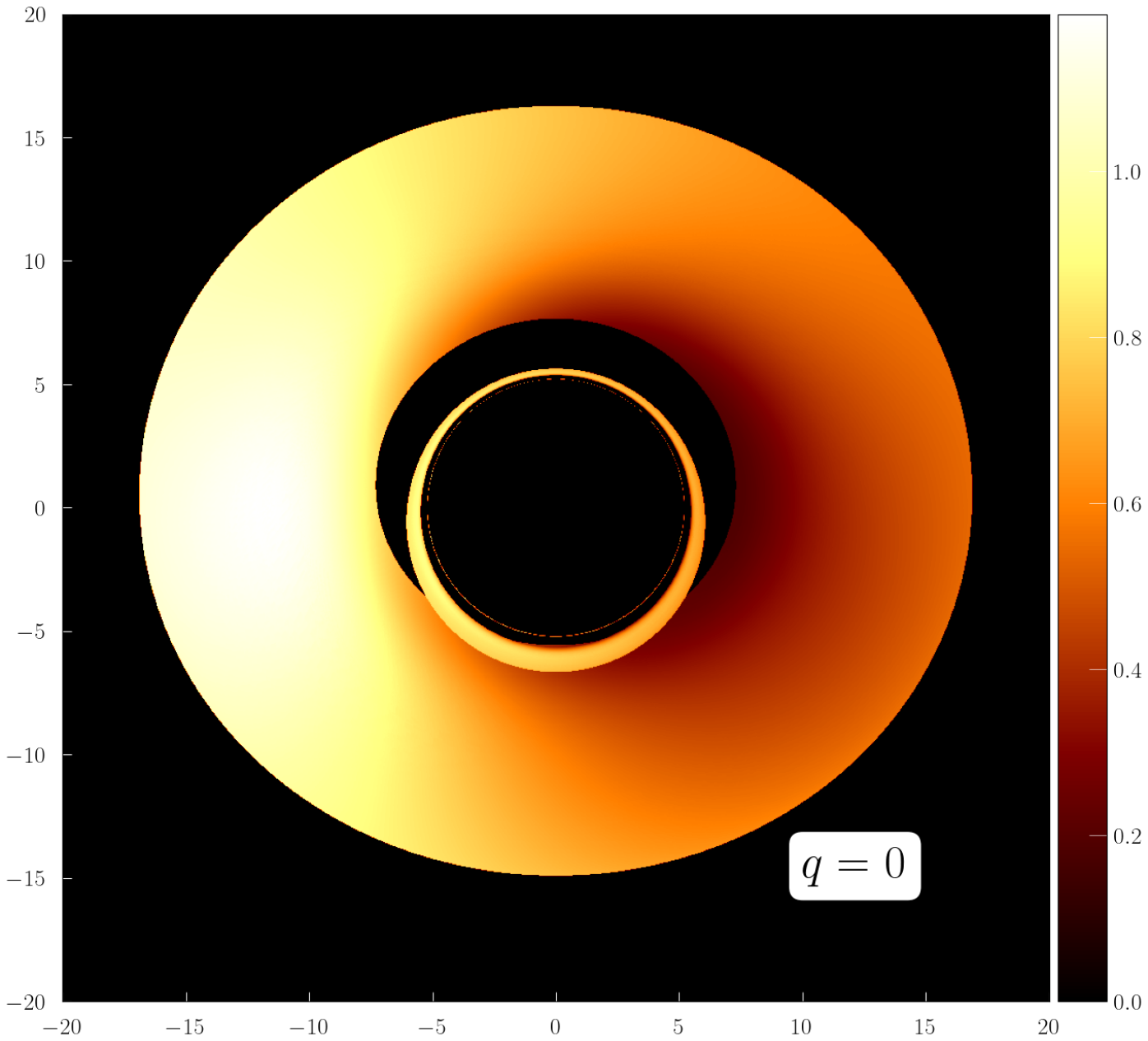}
		\caption{}
		\label{thin30q0}
	\end{subfigure}
	\begin{subfigure}[b]{0.48\textwidth}
		\includegraphics[width=\textwidth]{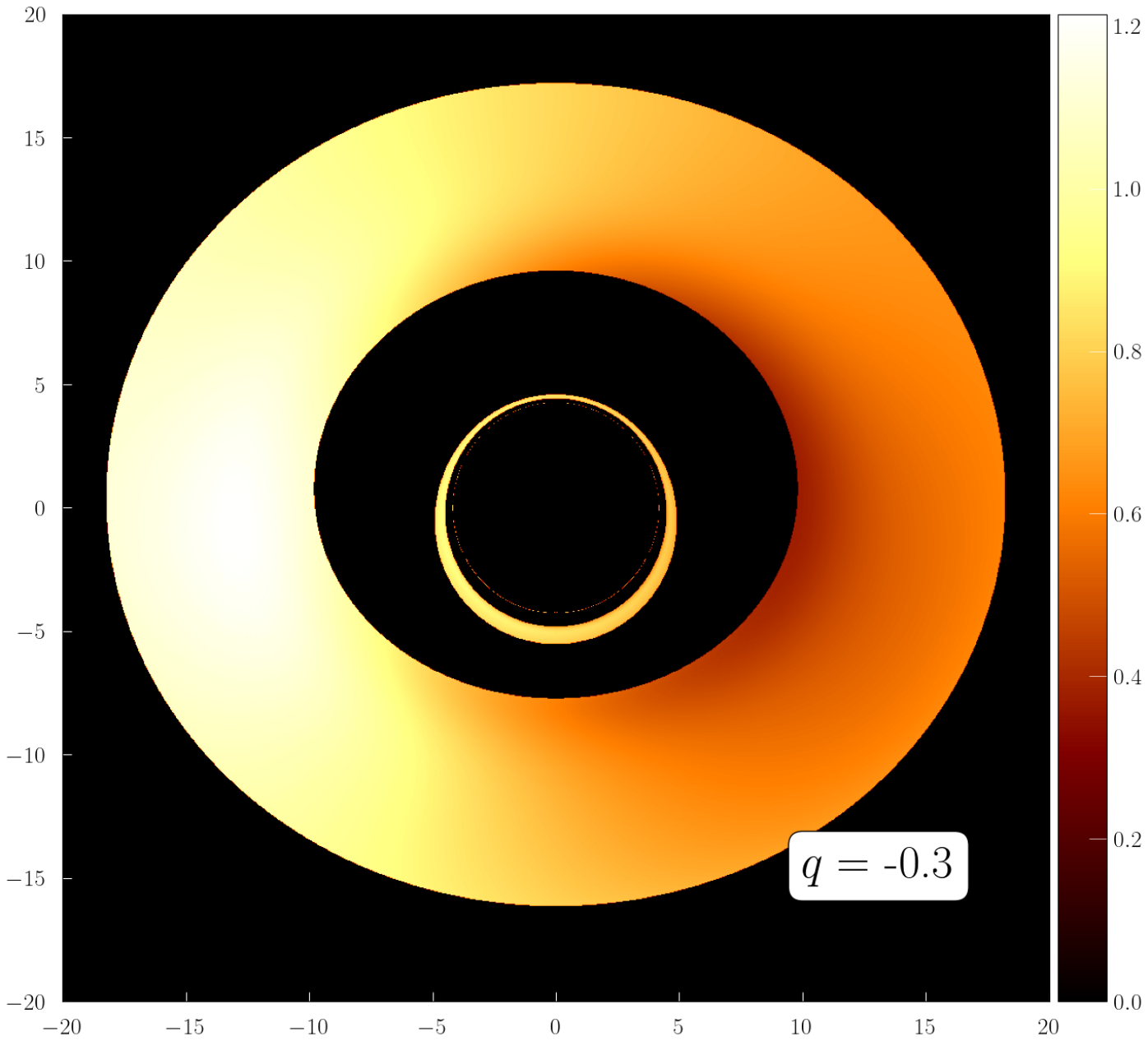}
		\caption{}
		\label{thin30q03}
	\end{subfigure}
	\caption{Optically thin magnetized tori around a naked singularity seen from $\theta=30$°. (a) $q=0$. (b) $q=-0.3$.}
	\label{thin30}
\end{figure*}

\begin{figure*}
	\centering
	\begin{subfigure}[b]{0.48\textwidth}
		\includegraphics[width=\textwidth]{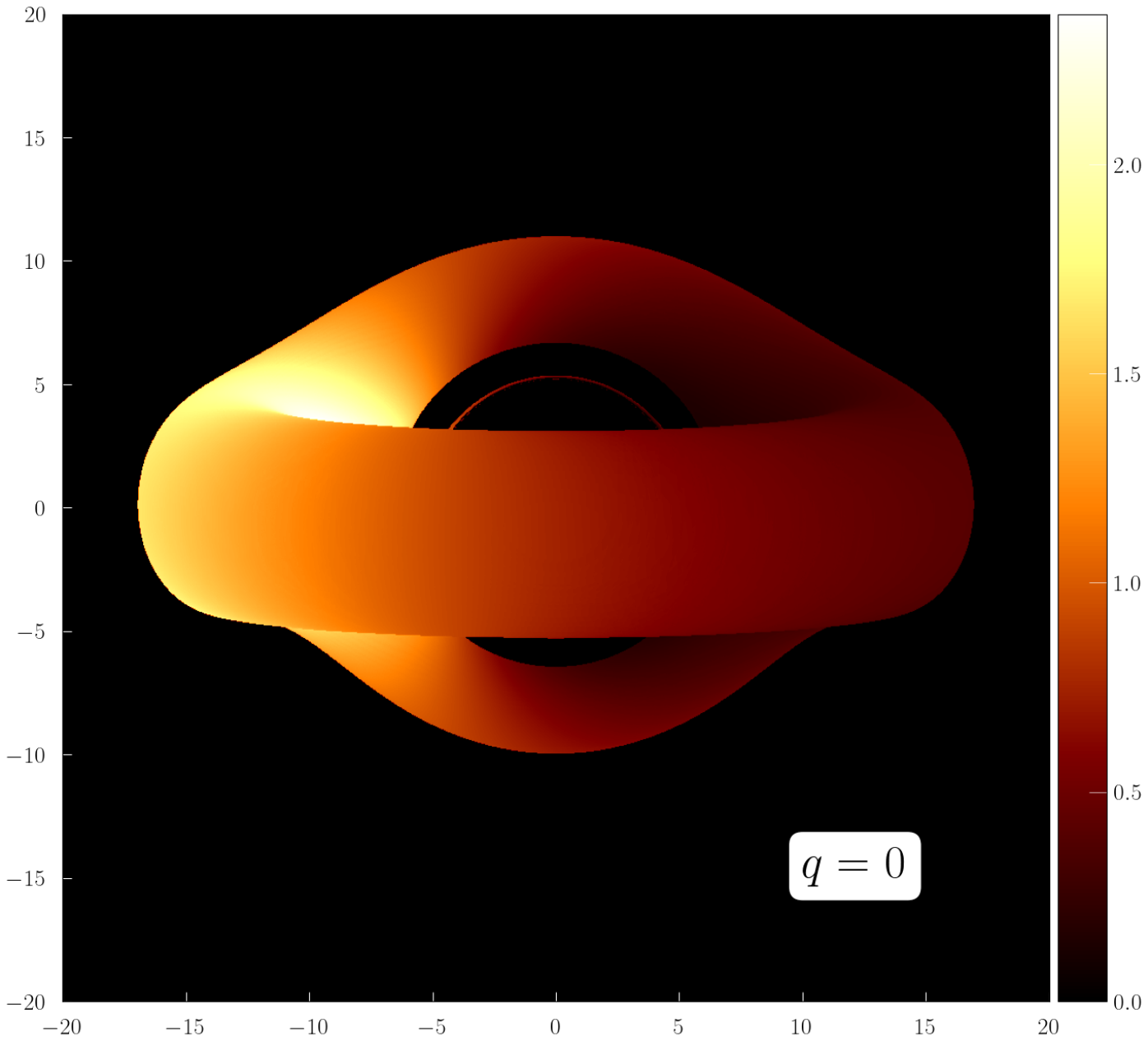}
		\caption{}
		\label{thick85q0}
	\end{subfigure}
	\begin{subfigure}[b]{0.48\textwidth}
		\includegraphics[width=\textwidth]{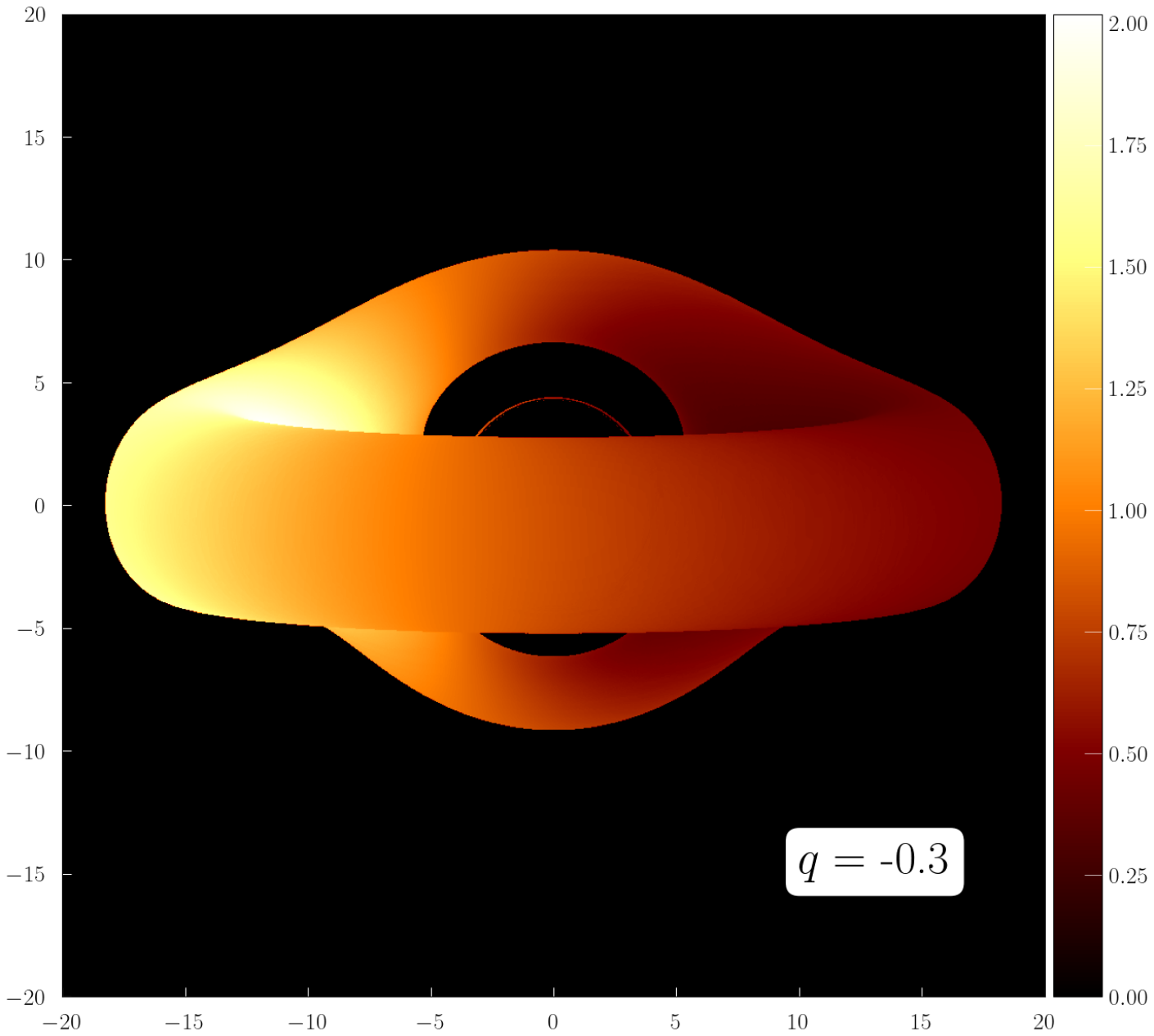}
		\caption{}
		\label{thick85q03}
	\end{subfigure}
	\caption{Optically thick magnetized tori around a naked singularity seen from $\theta=85$°. (a) $q=0$. (b) $q=-0.3$.}
	\label{thick85}
\end{figure*}

Finally, we present the results for the intensity maps of magnetized tori using a synchrotron radiation model in $q$-metric spacetime. For this section, a representative value of $\beta_c = 0.1$ was chosen. Figures \ref{thin85} and \ref{thin30} show the configurations of optically thin disks for observation angles of 85° and 30°, respectively. In Figure \ref{thick85}, we depict optically thick tori for an observation angle of 85°.  Through these panels,  we display the intensity maps corresponding to $q = -0.3$ and $q = 0$. We chose these representative values since the former clearly shows the impact of the quadrupole on the spacetime of the $q$-metric, while the latter serves as a control parameter by describing the case of Schwarzschild. Furthermore, in Figure \ref{thin85q0}, we observe a completely circular photon ring, a well-known characteristic of a static black hole \cite{luminet1979image}.

In Figure \ref{thin85q03} we notice that the photon ring has a more oval shape compared to the one observed in Figure \ref{thin85q0}. This effect was previously found for thin disks around naked singularities in \cite{shaikh2019can}, and we associate it with the deviation from the spherical symmetry of the compact object when $q= -0.3$.  Additionally, the disk surrounding a prolate object obtained for this work is more extended and located further away from the compact object, which is consistent with the results presented in Section \ref{ResI}, for the proposed mix of parameters. Figure \ref{thin85q03} also displays the typical features of the shadow of a compact object, including the gravitational redshift, visible as a bright area on the left side of the disk, and the gravitational lensing effect, which allows us to observe the opposite of the torus, located behind the singularity. Similarly, upon analyzing Figure \ref{thin30}, we can observe that the visible area of the disk increases, which is consistent with the increase in flux observed in the previous section when the observation angle was varied. 

\begin{figure*}
	\centering
	\begin{subfigure}[b]{0.48\textwidth}
		\includegraphics[width=\textwidth]{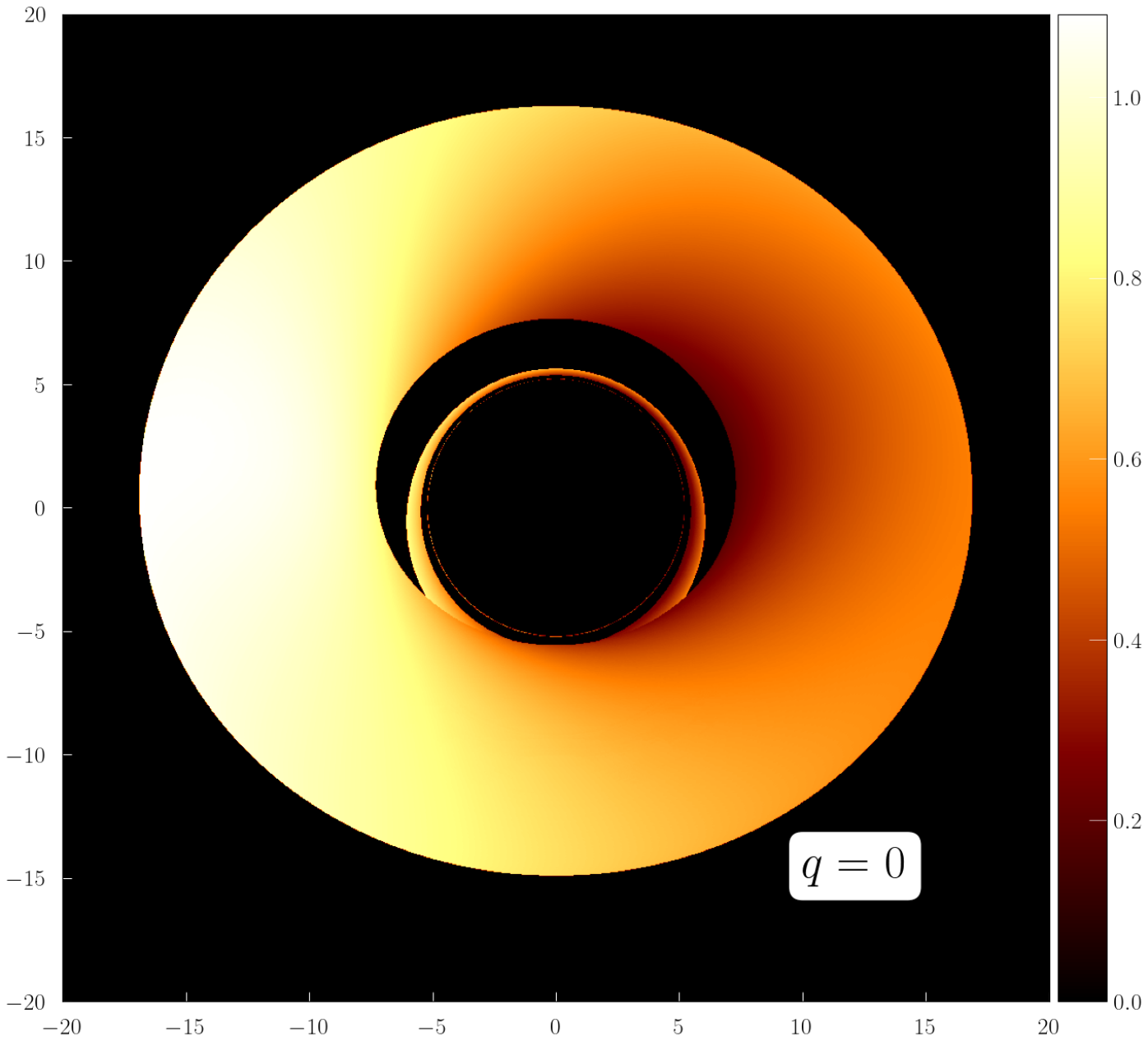}
		\caption{}
		\label{thick30q0}
	\end{subfigure}
	\begin{subfigure}[b]{0.48\textwidth}
		\includegraphics[width=\textwidth]{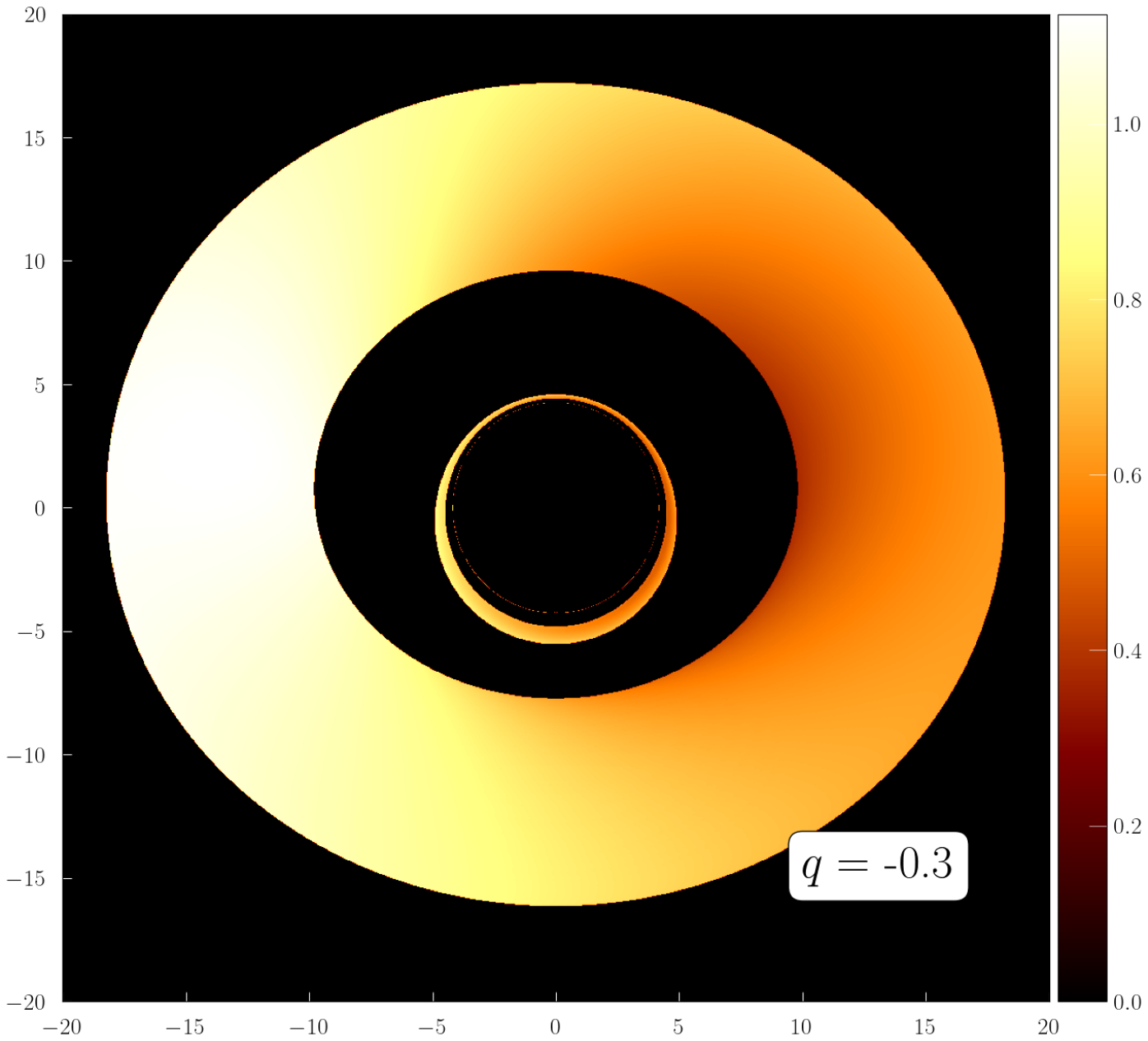}
		\caption{}
		\label{thick30q03}
	\end{subfigure}
	\caption{Optically thick magnetized tori around a naked singularity seen from $\theta=30$°. (a) $q=0$. (b) $q=-0.3$.}
	\label{thick30}
\end{figure*}

On the other hand, optically thick disks are shown in Figures \ref{thick85} and \ref{thick30} for observation angles of $\theta = 85$° and $\theta = 30$°, respectively. Generally, for this type of disk, the same effects of extension and separation from the compact object as reported for thin disks are observed. However, due to the nature of the optically thick disks, only the torus surface is observable. As a consequence, in Figure \ref{thick85}, we cannot see the complete shadow of the compact object. Furthermore, the self-eclipsing phenomenon becomes evident, whereby some regions of the torus surface obscure others.

Likewise, upon examining Figure \ref{thick30}, we can observe a morphology that is practically identical to that of the thin disks. However, a difference in the maximum intensity reached by these tori is noticeable, with the optically thick case showing a lower value, consistent with the previously presented model explanation.

\section{Discussion \& Conclusions} \label{Conclusion}

In this work, we revisited the developments of \cite{faraji2021magnetised} and systematically constructed our own family of magnetized tori in equilibrium around a naked singularity described by the $q$-metric. Our study is characterized by fixing the ADM mass of the spacetime, which results in configurations with different values of the angular momentum. We then simulated the intensity maps and emission spectra for both optically thick and thin disks. By doing so, we evaluated the impact of the $q$ parameter on the characterization of disks with a toroidal magnetic field and identified differences in observable quantities such as intensity and emission profiles, which will allow us to compare the theoretical model to the observations in the future. 

Firstly, we conducted a systematic study of tori equilibrium configurations in a spacetime with a fixed monopole $M_0$. Based on the distinct behaviour of the angular momenta corresponding to the orbits of interest for this case, we delimited a range of values  $\{ -0.3 \leq q \leq 0.2\}$ for the construction of this type of disk. In this case, we found that the configuration with the greatest extent was that for the value $q = -0.3$, which is also characterized by having the highest value of the angular momentum $l_0$. Furthermore, regarding the geometric features of the disks, we noticed that for configurations where there is a compact object with prolate deformation, $q < 0$, the disks take on a more rounded shape, while for the prolate case, $q > 0$, their shape resembles that of a seed. 
 
In addition, when compared between the different cases, we found that despite the extent of the disk, the maximum values of mass density were highly similar among them. As a result, this feature was associated predominantly with the value of the mass monopole $M_0$. Moreover, when analyzing the impact of the magnetization parameter we found that the variations of $\beta_c$ have a less considerable effect when the mass quadrupole does not vanish. Besides, we validate that the choice of the parameter $\beta_c$ does not affect the geometric properties of the disk but rather the mass distribution along it.

In the second part of the study, the focus was on analyzing the photons that manage to escape the gravitational pull of the compact object, allowing a distant observer to characterize the accretion disk in the vicinity of this astrophysical object. The results presented here constitute the first series of simulations of intensity maps and emission profiles for magnetized tori using the Komissarov formulation in the $q$-metric spacetime. To model the synchrotron radiation, a power law was assumed.

Then, upon analyzing the emission profiles, we found that for values of $q < 0$, which is the disk with the largest extension due to a combination of a negative value of q and a higher angular momentum $l_0$, the magnitude of the flux increases in comparison to the control case $q = 0$ for both optically thin and thick tori. Moreover, we found that upon increasing the degree of magnetization of the disk, the flux amplitude decreases, a behaviour consistent with the fact that the source function decreases as $B^{-1/2}$. For thick disks, the flux reaches higher values at an observation angle of 30° compared to 85° due to the self-eclipsing phenomena playing an important role for records near the equator. Overall, the results of the emission profiles of these disks are in agreement with previous findings in the literature, characterized by a prominent magnitude peak associated with blueshifted emission and a less pronounced peak related to redshifted emission.

On the other hand, we found larger flux values for thin disks than in the previous case, which is associated with a longer optical path followed by photons in this type of disk and the reduction in the magnitude of the self-eclipsing phenomenon due to the very nature of these configurations. Moreover, we note that a distinctive result when studying magnetized disks in this spacetime is the increase in the magnitude of the redshift peak for the case with $q = -0.3$, which even surpasses the height of the peak associated with the blueshifted emission. However, it is necessary to remark that this result can not be solely attributed to the $q-$ value since the high angular momentum of this configuration could also be playing a role in this behaviour.

Finally, concerning the intensity maps, we observed a change in the shape of the photon ring, which became more oval-shaped. This is attributed to the prolate nature of the compact object at the centre of the configuration. Additionally, we observed a greater distance between the disk's inner edge and the photon ring that defines the object's shadow, as well as an increase in intensity compared to the reference case with $q = 0$. We also observed the expected effects of this type of astrophysical system, such as gravitational redshift and lensing. It is worth noting that variations in the observation angle or the optical properties of the disk result in changes in the visible section of the disk and the impact of self-eclipsing effects, respectively.

Thus, we conclude that the $q$ parameter plays a relevant role in the study of the emission spectra of geometrically thick disks with a toroidal magnetic field in the $q$-metric spacetime, exhibiting a set of unique properties that may prove useful for comparing observations of a possible black hole or one of its imitators described by this metric. Nonetheless, further analysis is required to single out the contribution to these results due to the $q$ parameter from the role of the angular momentum of the disk $l_0$.

The next step in this study is to use the solution of thick disks around the $q$-metric as a starting point for dynamic simulations of the accretion flow, to more accurately evaluate the physical viability of a naked singularity as a realistic black hole mimicker. 

\section*{Data availability statement}
The data that support the findings of this study are available upon request from the authors.

\section*{Acknowledgements}

F.D.L-C was supported by the Vicerrectoría de Investigación y Extensión - Universidad Industrial de Santander, under Grant No. 3703.

\section*{References}

\providecommand{\newblock}{}

\end{document}